%% file: main.tex
\documentclass[10pt,twocolumn,letterpaper]{article}

\usepackage[pagenumbers]{cvpr} %

\input{preamble}

\definecolor{cvprblue}{rgb}{0.21,0.49,0.74}
\usepackage[pagebackref,breaklinks,colorlinks,allcolors=cvprblue]{hyperref}

\def\confName{CVPR}
\def\confYear{2025}

\title{\ours: Fine-Grained Monocular Non-rigid  3D Surface Tracking \\
with Neural Deformation Fields}

\author{Navami Kairanda\textsuperscript{1}\; 
Marc Habermann\textsuperscript{1, 2}\; 
Shanthika Naik\textsuperscript{3}\; 
Christian Theobalt\textsuperscript{1, 2}\; 
Vladislav Golyanik\textsuperscript{1}\\
\vspace{12pt}
\textsuperscript{1}MPI for Informatics, SIC\quad\,
\textsuperscript{2}VIA Research Center\quad\,
\textsuperscript{3}IIT Jodhpur}

\begin{document}
\maketitle
\input{sections/0_abstract}    
\input{sections/1_introduction}
\input{sections/2_related_work}

\input{sections/3_method}
\input{sections/4_results}

\input{sections/5_conclusion}
\input{sections/7_acknowledgements}

{
    \small
    \bibliographystyle{ieeenat_fullname}
    \bibliography{main}
}
\input{sections/6_appendix}

\end{document}

%% file: preamble.tex
\usepackage{comment}
\usepackage{csquotes}
\usepackage{adjustbox}
\usepackage{wrapfig}
\usepackage{tabularx}
\usepackage{siunitx}
\usepackage[accsupp]{axessibility}  %
\usepackage{multirow}
\usepackage{titletoc}
\usepackage{tocloft}

\definecolor{first}{RGB}{255,178,178}
\definecolor{second}{RGB}{255,217,178}
\definecolor{third}{RGB}{255,255,178} 

\def\ours{Thin-Shell-SfT}

\def\phisft{$\boldsymbol{\phi}$-SfT}

\usepackage{soul}

%% file: sections/0_abstract.tex
\begin{abstract} 
3D reconstruction of highly deformable surfaces (e.g. cloths)
from monocular RGB videos is a challenging problem, and no solution provides a consistent and accurate recovery of fine-grained surface details. 
To account for the ill-posed nature of the setting, 
existing methods use deformation models with statistical, neural, or physical priors. 
They also predominantly rely on nonadaptive discrete surface representations (e.g.~polygonal meshes), perform frame-by-frame optimisation leading to error propagation, 
and suffer from poor gradients of the mesh-based differentiable renderers. 
Consequently, fine surface details such as cloth wrinkles are often not recovered with the desired accuracy. 
In response to these limitations, we propose \emph{\ours}, a new method for non-rigid 3D 
tracking that represents a surface as an implicit and continuous
spatiotemporal neural field. 
We incorporate continuous thin shell %
physics prior based on the Kirchhoff-Love model for spatial regularisation, which starkly contrasts the discretised alternatives of earlier works. %
Lastly, we 
leverage 3D Gaussian splatting to differentiably render the surface into image space and optimise the deformations based on analysis-by-synthesis principles. %
Our \emph{\ours} 
outperforms prior works %
qualitatively and quantitatively
thanks to our continuous surface formulation in conjunction with a specially tailored simulation prior and 
surface-induced 3D Gaussians.
See our project page at \url{https://4dqv.mpi-inf.mpg.de/ThinShellSfT}.
\end{abstract}

%% file: sections/1_introduction.tex
\section{Introduction} \label{sec:intro} 
\input{figures/fig_teaser} 
Non-rigid 3D reconstruction and tracking of general deformable surfaces from a monocular RGB camera is an important, challenging and ill-posed problem that is far from being solved~\cite{tretschk2023state}. 
It has applications in game development, robotics and augmented reality, to name a few areas. 
\par
Prior works on temporally-coherent general surface reconstruction can be grouped into \emph{Shape-from-Template} (SfT)~\cite{Bartoli2015, kairanda2022f, casillas2021isowarp} and \emph{Non-Rigid Structure-from-Motion}~\cite{Parashar_2020_CVPR, Sidhu2020, kumar2019jumping}; 
they rely on 
2D point tracks across monocular images (NRSfM)~\cite{Parashar_2020_CVPR, Sidhu2020} or between the input image and the template (SfT)~\cite{Bartoli2015, casillas2021isowarp}. 
Recent physics-based SfT approaches~\cite{kairanda2022f, stotko2023physics} demonstrate state-of-the-art results and cause a paradigm shift from geometric- \cite{Bartoli2015, casillas2021isowarp, Parashar2015} to physics-based constraints and from 2D-point-based~\cite{Ngo2015,casillas2021isowarp} to dense photometric loss using differentiable renderers. 
However, even such approaches as \phisft~\cite{kairanda2022f} 
do not support fine wrinkles and require multiple hours to reconstruct a limited number of frames ($\approx$50) of a single object, 
which is due to 
the underlying surface mesh representation, \ie,~an explicit and discrete. 
\textit{First,} determining the mesh resolution for a specific scene is difficult, \eg selecting a high resolution could prohibitively increase the memory and computational cost, 
while a lower resolution might not account for fine-scale deformations. 
\textit{Second,} the Finite Element Mesh (FEM)-based differentiable physics simulators~\cite{Liang2019,li2022diffcloth} lead to~\emph{inconsistent} simulations at different resolutions, preventing the adoption of coarse-to-fine strategies for tracking. 
\textit{Third,} simulator-based approaches~\cite{kairanda2022f, stotko2023physics} optimise frame-by-frame 
and can get stuck in local minima. 
\textit{Moreover,} mesh-based differentiable renderers used in recent approaches~\cite{kairanda2022f, stotko2023physics} are nonadaptive and do not support the complex remeshing required for the dynamic details arising in deformable scenes. 
\par
To overcome these limitations, our key idea is to replace meshes with an adaptive and continuous surface and deformation representation\footnote{Most prior methods use a discrete number $N$ of points or mesh vertices (including neural ones \eg \cite{Sidhu2020}) where $N$ %
needs to be decided beforehand and cannot be changed during optimisation, unlike our continuous formulation, allowing arbitrary queries.}. %
We tightly couple it with differentiable physics to guide the deformations while ensuring photometric consistency with the monocular images through 
analysis-by-synthesis. 
We model surfaces as Kirchhoff-Love thin shells~\cite{love1927treatise} and propose a physics-based \emph{continuous}  deformation prior. 
Our regulariser minimises the internal hyperelastic energy of the tracked surface, ensuring its physical plausibility and providing prior for occlusion handling. 
Unlike discrete priors operating on mesh vertices \cite{Yu2015,kairanda2022f}, %
our continuous prior updates any point on the surface, enabling physics supervision for fine-grained details such as wrinkles. 
While similar modelling has been applied to cloth simulation~\cite{kairanda2023neuralclothsim}, %
it has not been shown for inverse problems like monocular %
surface tracking, which
involves many unknowns (\eg, material, forces, contacts) and ambiguities between them. 
In addition to continuous prior, we perform joint space-time optimisation while taking the causality 
of deformation into account to impose temporal coherence (\ie~the current deformed state can update previous state parameters but not the future ones). 
As the differentiable renderer, we employ 3D Gaussian Splatting~\cite{kerbl20233d}, which recently emerged as a prominent technique for radiance field rendering. %
It integrates straightforwardly with our continuous per-point deformation model and offers high-quality image gradients due to the continuous volumetric radiance field formulation. 
We leverage differential geometric quantities from the thin shell physics to couple Gaussians to the surface and optimise the parameters of dynamically tracked Gaussians. 
While there are extensions of Gaussian splatting focussing on dynamic view synthesis~\cite{yang2023deformable, duan20244d} or multi-view static surface reconstruction \cite{waczynska2024games,guedon2024sugar}, we show how to adapt it 
for surface tracking from monocular videos with 
\emph{de-facto absent multi-view 3D reconstruction cues}~\cite{gao2022monocular}
such as the \phisft~\cite{kairanda2022f} dataset.  
The technical contributions of this paper are as follows: 
\begin{itemize} 
\item \ours, \ie a new method for monocular non-rigid 3D surface tracking operating on a continuous adaptive spatiotemporal representation;  
\item 
A continuous deformation prior based on the principles of the Kirchhoff-Love thin shell theory and application of 
such a prior in an inverse problem (shape from template); 
\item %
Adaptation of 3D Gaussian Splatting for 3D tracking of highly deformable dynamic surfaces forming folds and fine wrinkles, captured by a static monocular camera. 
\end{itemize} 
Our experimental results on the challenging \phisft~benchmark \cite{kairanda2022f} show a significant improvement over the state of the art in terms of reconstruction accuracy; see Fig.~\ref{fig:teaser}. 

%% file: figures/fig_teaser.tex
\begin{figure}
\centering
\includegraphics[width=\linewidth]{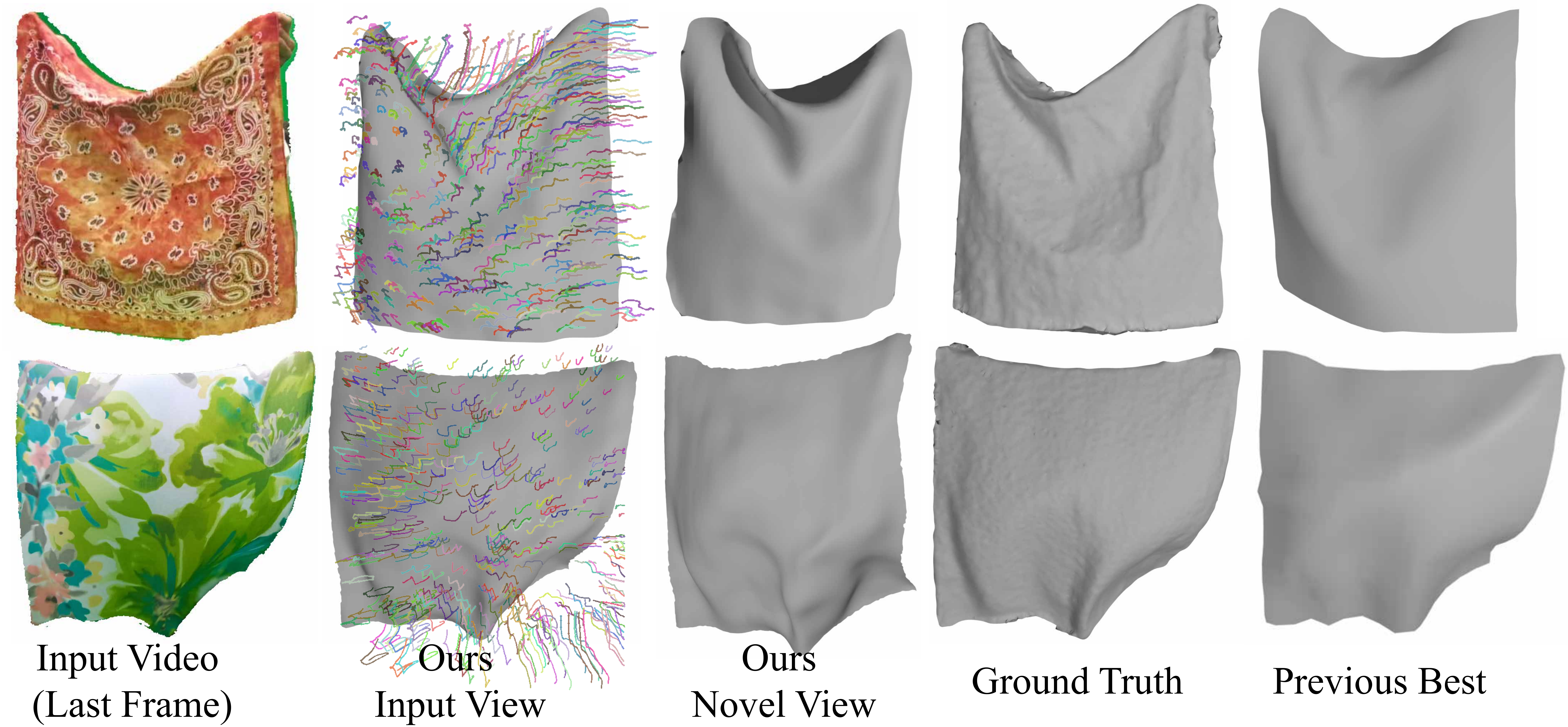}
  \caption{ 
Our \emph{\ours} approach reconstructs high-fidelity deformable 3D surface geometry with fine-grained wrinkles from a monocular video, while the previous best method \cite{kairanda2022f} struggles. The coloured tracks (``Ours, Input View'') visualise the 3D Gaussian trajectories 
over time and across all input video frames. 
  }
  \vspace{-18pt}
  \label{fig:teaser}
\end{figure}

%% file: sections/2_related_work.tex
\section{Related Work} 
\label{sec:related}
We review methods for monocular reconstruction of general non-rigid surfaces. 
They differ in their assumptions about the deformation model, data terms, and the priors; 
we refer to a recent survey by Tretschk~\etal~\cite{tretschk2023state} for an in-depth review. 
Similar to ours, many recent works integrate Gaussians with physics or clothing but in very different contexts such as simulation~\cite{xie2024physgaussian,feng2024splashing}, material estimation~\cite{zhang2024physdreamer,liu2024physics3d}, and tracking from multi-view video~\cite{longhinicloth,duisterhof2023md,PhysAavatar24}.

\noindent\textbf{Shape-from-Template (SfT)}
methods~\cite{Salzmann2007, Perriollat2011, Ngo2015, Yu2015, Pumarola2018, FuentesJimenez2021, kairanda2022f, haouchine2017template} assume a single static shape or \emph{template} as a prior. 
A template often corresponds to the first frame of the sequence~\cite{Ngo2015, Bartoli2015, kairanda2022f} and, in other cases, to the rigid initialisation~\cite{fuentes2022deep, Shimada2019}. 
SfT approaches~\cite{Bartoli2015, Yu2015, Ngo2015, NRST_GCPR2018} employ a 3D-2D reprojection constraint and deform the template using temporal and geometric soft constraints to encourage physical plausibility. 
On the other hand, learning-based SfTs~\cite{Golyanik2018, Pumarola2018, Shimada2019, FuentesJimenez2021} encode the template in the network and regress the surface deformations. 
They are sometimes object-specific with template encoded in the neural network weights~\cite{fuentes2022deep} and often object-generic~\cite{FuentesJimenez2021, Golyanik_2019,Shimada2019} when supervised with synthetic datasets. 
Recently, Kairanda~\etal\cite{kairanda2022f}~and Stoko~\etal~\cite{stotko2023physics} explain 2D observations through physics-based simulation of the deformation process and employ differentiable rendering for per-sequence gradient-based optimisation. 
In contrast to traditional SfT~\cite{Bartoli2015, Parashar2015, casillas2021isowarp}, they do not require registrations (the correspondence between the 3D template and the image). 
Ours falls under this category as we require the template corresponding to the first frame. 
In contrast to the prior works, ours is the first method to continuously represent the surface and its dynamics with a neural field capable of representing high-frequency signals~\cite{sitzmann2020implicit}.
We supervise with thin shell physics constraint applied at continuous points in the domain and employ the Gaussian Splatting functionality~\cite{kerbl20233d}. 

\noindent\textbf{Non-Rigid Structure from Motion (NRSfM)}
\cite{Bregler2000, Torresani2008, Akhter2008, Paladini2009, Dai2014} relies on motion cues of 2D point tracks over the input monocular images
and outputs per-frame camera-object poses and the 3D shapes. 
Recent \emph{dense} NRSfM methods \cite{Sidhu2020, Parashar_2020_CVPR, Ansari2017, Kumar2018, Golyanik_2019, Wang2022CVPR, Grasshof2022} rely on per-pixel multi-frame optical flow or video registration \cite{Garg2013MFOF}. 
In contrast, we do not make a restrictive assumption of available 2D point tracks and 
track highly challenging cloth deformations (fine wrinkles), which goes beyond NRSfM capabilities. 

 \noindent \textbf{Dynamic Novel View Synthesis.}
Radiance field methods~\cite{mildenhall2020nerf, wang2021neus, Mueller2022, kerbl20233d} learn volumetric scene representation %
for high-quality novel-view synthesis
from multi-view images,  
which are extended to support 
dynamic scenes~\cite{tretschk2021nonrigid, park2021hypernerf, wang2022neus2, fridovich2023k, yang2023deformable, das2024neural, johnson2023unbiased,kratimenos2024dynmf}. 
Though they demonstrate impressive results, most do not show or evaluate geometry reconstruction. 
Moreover, they require multi-view cues \cite{Pumarola2018, gao2022monocular}; in contrast, we focus on dynamic scenes captured from a single static camera without such cues. 

\par 

\noindent\textbf{Physics-based Priors} 
have been applied in inverse problems~\cite{PhysCapTOG2020, Rempe2020, Li2021, Guo2021}. 
The examples range from 3D human pose estimation~\cite{PhysAwareTOG2021, Li2021} and parameter estimation~\cite{Liang2019, Weiss2020CorrespondenceFreeMR, murthy2021gradsim, hofherr2023neural} to dense SfT~\cite{kairanda2022f, stotko2023physics, DecafTOG2023}. 
Recent works~\cite{Chen_2022_CVPR, qiao2022neuphysics} extend neural 
representations to physical parameter inference from videos. 
Some methods~\cite{yang2023ppr, kandukuri2020_diffphys, Jaques2020Physics-as-Inverse-Graphics:} combine differentiable simulation and rendering. 
In the inverse setting, 
we are the first to impose \emph{continuous} physics constraints 
while prior methods use mesh-based simulators.

%% file: sections/3_method.tex
\section{Preliminaries}
\label{ssec:preliminaries}
We review 3D Gaussian Splatting~\cite{kerbl20233d} and NeuralClothSim~\cite{kairanda2023neuralclothsim}, from which we take inspiration for the 
data term (ensuring consistency of reconstructions with input images) and the prior term (encouraging physical plausibility). 

\noindent\textbf{3D Gaussian Splatting (3DGS)} 
models a scene as a volumetric radiance field with a dense set of 3D Gaussians, each defined by its position (mean), anisotropic covariance, opacity, and colour. 
$N_g$ Gaussians are represented as
\begin{equation}
    \mathcal{G} = \{ \mathcal{N}(\mathbf{x}_i, \mathbf{R}_i, \mathbf{S}_i), o_i, \mathbf{c}_i\}_{i=1}^{N_g}, %
    \label{eq:gaussians}
\end{equation}
where $\mathbf{x}$ denotes the position;  $\mathbf{R S S^\top R^\top}$ is the anisotropic covariance parameterised as an ellipsoid with
scale $\mathbf{S}$ and rotation $\mathbf{R}$; $o$ is the opacity, and $\mathbf{c}$ is the colour of the Gaussian. 
In 3DGS~\cite{kerbl20233d}, $\mathcal{G}$ is optimised through gradient-based training with multi-view image loss. 
Learning a 3DGS representation from monocular inputs requires substantial adjustments in different contexts \cite{das2024neural,yang2023deformable}, and we show its utility in SfT for the first time.

\noindent\textbf{NeuralClothSim}~\cite{kairanda2023neuralclothsim} is a recent quasistatic cloth simulator representing surface deformations as a coordinate-based implicit neural deformation field (NDF). %
Given a target simulation scenario specified by the initial surface state, material properties and external forces, NeuralClothSim learns the equilibrium deformation field using the laws of the Kirchhoff-Love thin shell theory. 
Upon convergence, the equilibrium state can be queried continuously and consistently at multiple resolutions. %
These properties, combined with the memory adaptivity of neural fields, make NeuralClothSim well-suitable for inverse problems like ours. 

Kairanda \etal~\cite{kairanda2023neuralclothsim} model a cloth quasistatically as an NDF 
$\mathbf{u}(\boldsymbol{\xi}):\Omega \mapsto \mathbb{R}^3$ defined on its curvilinear coordinate space $\Omega$. 
For a volumetric thin shell such as cloth, the Kirchhoff-Love model~\cite{wempner2003mechanics} offers a reduced kinematic parameterisation of the volume characterised by a 2D \emph{midsurface} that fully determines the strain components throughout the thickness.
Following thin shell assumptions, they decompose the Green strain due to deformation $\mathbf{u}$---parameterised by a multilayer perceptron (MLP)---into membrane strain $\boldsymbol \varepsilon$ and bending strain $\boldsymbol \kappa$, measuring the in-plane stretching and the change in curvature, respectively. 
NeuralClothSim then computes the internal hyperelastic energy $\Psi[\boldsymbol \varepsilon, \boldsymbol \kappa;\boldsymbol{\Phi}]$ as a functional of the geometric strains and the cloth's material properties $\boldsymbol{\Phi}$. %
Under the action of external force $\mathbf{f}$, their neural solver
finds the equilibrium solution following the principle of minimum potential energy, where the potential $\int_{\Omega}\Psi \, d\Omega - \int_{\Omega} \mathbf{f}\cdot\mathbf{u} \, d\Omega$ is employed as the loss function. 
Note that the forward model of NeuralClothSim \cite{kairanda2023neuralclothsim} cannot be naively applied or extended (e.g.,~when coupled with a differentiable renderer, estimating physics parameters with the loss of NeuralClothSim would not converge due to the lack of lower bound) in our inverse setting.

\section{Method} 
\label{sec:method}
\input{figures/fig_overview}
We propose \ours, a new template-based method for the fine-grained 3D reconstruction of a deforming surface (such as cloth, paper or metal) seen in a monocular RGB video; see \cref{fig:overview}. %
We aim to reconstruct continuous 3D surfaces $\{\mathbf{S}_t\}_{t\in [1,\ldots,T]}$ corresponding to the input image sequence $\{\mathbf{I}_t\}_{t}$ and optional masks $\{\mathbf{M}_t\}_{t}$. %
Similar to previous surface tracking methods \cite{Ngo2015, kairanda2022f}, we assume a static camera with known calibration and take the surface $\mathbf{S}_1$ corresponding to the first frame $t{=}1$ as a template. 
\par

In contrast to the discrete representations (\eg points, meshes) used in previous monocular tracking approaches, 
our deformation model encodes the surface and its dynamics as continuous and adaptive neural fields (\cref{ssec:deformation_model}). 
We optimise the neural model
by relating input monocular views to the estimated surface states using  3DGS~\cite{kerbl20233d}, %
where the Gaussians are initialised and dynamically tracked with parameters induced from the surface deformation (\cref{ssec:image_loss}). 
To ensure plausible deformations %
 due to the inherent ambiguity of the monocular setting and in contrast to prior discrete regularisers, we model the surfaces as thin shells imposing the continuous Kirchhoff-Love physics constraints~\cite{kairanda2023neuralclothsim} on the neural field, allowing us to reconstruct fine-grained details (\cref{ssec:physics_loss}). 
\subsection{Deformation Model and Parameterisation} %
\label{ssec:deformation_model}
\paragraph{Deformation Model.} 
As our deformation model, we employ a continuous surface representation and deformation dynamics modelled as coordinated-based neural fields. 
Consider a surface with initial state $\mathbf{S}_1 \subset \mathbb{R}^3$ and 2D parameterisation $\Omega \subset \mathbb{R}^2$.
For any parametric point $\boldsymbol{\xi} := (\xi^1, \xi^2) \in \Omega$, %
we represent its position on the initial surface state with a mapping  $\mathbf{\bar x}(\boldsymbol{\xi} ): \Omega \to \mathbf{S}_1$.
Furthermore, we encode the time-varying spatial location of $\boldsymbol{\xi}$ on the tracked surface $\mathbf{S}_t$ as $\mathbf{x}(\boldsymbol{\xi}, t): \Omega \times [1,...,T]\to \mathbf{S}_t$, with 
\begin{equation}
\small 
\begin{split}
        \mathbf{x}(\boldsymbol{\xi}, t) &= \mathbf{\bar{x}}(\boldsymbol{\xi}) + \mathbf{u}(\boldsymbol{\xi}, t), \;
        \text{ with }\;
        \mathbf{u}(\boldsymbol{\xi}, 1)=0\;\text{ and } 
        \\
        \mathbf{u}(\boldsymbol{\xi}, t) &= \lambda \mathbf{u}(\boldsymbol{\xi}, t-1) + \mathcal{F}(\boldsymbol{\xi}, t), \forall t>1, 
        \label{eq:surface_point}
\end{split}
\end{equation}
where $\mathbf{u}(\boldsymbol{\xi}, t)$ is the deformation field, 
$\lambda>0$ is a scalar value and $\mathcal{F}(\boldsymbol{\xi}, t)$ represents the estimated deformation offset. %
Here, we encourage the conservation of momentum by setting the deformation of the future surface states in the direction of the previous deformations. %

\noindent\textbf{Parameterisation of the Template and Deformations.} 
Next, we present an adaptive parameterisation of the above deformation model that allocates capacity to dynamic deformation details. %
Assuming that the template is provided as a %
mesh $\mathcal{M} \subset \mathbf{S}_1$, 
we generate the corresponding 2D parametric space $\mathcal{T} \subset \Omega$ using established techniques (\eg~conformal parameterisation~\cite{libigl}). 
To learn a continuous representation of the initial state, we fit an MLP, which we call \emph{neural reference field} (NRF) $\mathbf{\bar x}(\boldsymbol{\xi};\Upsilon)$ to the template mesh parameterisation $\mathcal{T} \mapsto \mathcal{M}$. 
For training NRF, we construct points in the parametric and vertex space with randomly sampled barycentric coordinates and train with $\ell_1$ geometry loss. 
Akin to NRF, we regress the spatio-temporal \emph{neural deformation field} (NDF) $\mathbf{u}(\boldsymbol{\xi}, t;\Theta)$ using another MLP predicting deformation offsets, $\mathcal{F} (\boldsymbol{\xi}, t;\Theta) : \Omega \times [1,...,T] \to \mathbb{R}^3 $. 
Before training, the initial estimate for tracked surface points 
$\mathbf{x}(\boldsymbol{\xi}, t;\Theta)$  is the sum of noisy NDF deformations, $\mathbf{u}(\boldsymbol{\xi}, t;\Theta)$ and the pre-trained NRF $\mathbf{\bar{x}}(\boldsymbol{\xi};\Upsilon)$ as given by \cref{eq:surface_point}. %
Our key idea is to optimise the NDF weights $\Theta$ to ensure photometric consistency of the tracked surfaces with input monocular views while minimising the internal energy of the surface states modelled as a thin shell.
The NRF pre-fit enables the coupling of Gaussians to the surface and computing surface metrics for the thin shell energy. 
We must represent high-frequency signals, such as fine folds and wrinkles on the tracked surface. 
Moreover, our physics loss requires the computation of higher-order derivatives. 
Hence, we use sine 
activation~\cite{sitzmann2020implicit} in both MLPs. %

\noindent\textbf{Inference.}
At testing, $\mathbf{x}(\boldsymbol{\xi}, t;\Theta)$ provides \textit{continuous} access to the reconstructed surface, where the NRF and NDF networks can be consistently queried at 
varied spatio-temporal resolutions. 
Moreover, our surface deformation model provides temporal correspondences.
Meshing and texture mapping can be achieved in the parametric domain and then transferred to the tracked surface in 3D using \cref{eq:surface_point}. 
We next describe how to optimise the deformation field $\mathbf{u}(\boldsymbol{\xi}, t;\Theta)$ that ensures consistency with monocular images and the physical plausibility of reconstructions. 
\subsection{Surface Tracking with Gaussian Splatting} %
\label{ssec:image_loss}
Recall that we have an NDF parameterised by $\Theta$ that outputs the deformation $\mathbf{u}(\boldsymbol{\xi}, t;\Theta)$ at any time step. 
We seek to optimise the NDF so that the tracked surfaces $\{\mathbf{S}_t\}_t$ (evaluated with \cref{eq:surface_point}) 
generate images matching with input views $\{\mathbf{I}_t\}_t$.
A straightforward approach to encourage 3D-to-2D consistency is to minimise the photometric loss with differentiable rendering.
We choose Gaussian Splatting~\cite{kerbl20233d} as the differentiable renderer, as it seamlessly integrates with our per-point (\cref{eq:surface_point}) continuous deformation model. 

\noindent\textbf{Initialisation of Gaussians.}
To initialise a set of Gaussians, we sample $N_{g}$ well-distributed points (\eg Poisson disk sampling~\cite{bridson2007fast}) from the surface of the input template mesh $\mathcal{M}$.
The sampled points include parametric coordinates $\{\boldsymbol{\xi}_i\}_{i\in [1,\ldots,N_g]}$, their corresponding positions $\{\mathbf{\bar x}_i\}_i$, %
and colours $\{\mathbf{c}_i\}_i$.
Next, we will motivate our approach to parameterising Gaussian rotations and scales. 
Unlike the original multi-view~\cite{kerbl20233d} or dynamic Gaussian methods~\cite{yang2023deformable, duan20244d} 
that leverage multi-view cues~\cite{gao2022monocular}, we have a single static camera and, consequently, a challenging monocular setup. 
Thus, learning the anisotropic covariance of 3D Gaussians from all input frames can lead to poor results due to the inherent monocular ambiguities between deformation, texture and appearance. 
In our datasets, the initial surface state is reasonably (but not exactly) flat and fully visible. %
With this assumption, we take the initial surface normal as a proxy for the viewing direction and then fix the scale along normal while allowing for optimisation of the scales along the initial surface tangents. 
More concretely, we set the rotation matrix of $i$-th Gaussian equal to  %
the local basis vectors, \ie $\mathbf{R}_i \equiv [\mathbf{\bar a}_1 \; \mathbf{\bar a}_2 \; \mathbf{\bar a}_3]_i$ (see \cref{ssec:physics_loss} for computation) of the initial state (NRF) $\mathbf{\bar x}(\boldsymbol{\xi}_i;\Upsilon)$ %
and the scale $s_3$ as small $\epsilon$ along surface normal $\mathbf{\bar a}_3$. 
Finally, the dynamically tracked surface-induced  Gaussian $\{\mathcal{G}_t\}_t$ can be written as 
\begin{equation}
\begin{aligned}
     \mathcal{G}_t &= \{\mathcal{N}(\mathbf{x}(\boldsymbol{\xi}_i, t; \Upsilon, \Theta), 
     [\mathbf{\bar a}_1(\boldsymbol{\xi}_i; \Upsilon), \mathbf{\bar a}_2(\boldsymbol{\xi}_i; \Upsilon), 
     \\
     &\mathbf{\bar a}_3(\boldsymbol{\xi}_i; \Upsilon)], 
     (s_1, s_2, \epsilon)_i),  o_i, \mathbf{c}_i\}_{i=1}^{N_g}
\end{aligned}
\label{eq:tracked_Gaussian}
\end{equation}
following \cref{eq:gaussians}. 
Note that the deformed positions $\mathbf{x} (\boldsymbol{\xi}_i, t)$ are computed using our deformation model (\cref{eq:surface_point}) and all other Gaussian parameters are shared across surface states.

\noindent\textbf{Optimisation of NDF and Gaussian Parameters.} 
For optimisation, we use the $\ell_1$ photometric loss similar to 3DGS~\cite{kerbl20233d}. %
If segmentation masks $\{\mathbf{M}_t\}_t$ are available, an additional silhouette loss is added to primarily speed up training. %
Thus, our data loss reads as 
\begin{equation}
\begin{split}
    \small
    \mathcal{L}_{d}(s_1, s_2, &o, \mathbf{c}, \Theta) = \mathcal{R} (\mathcal{G}_1, \mathbf{I}_1; s_1, s_2, o, \mathbf{c}) + \\ %
    & \sum_{t=2}^{T} \mathcal{R} (\mathcal{G}_t, \mathbf{I}_t;\Theta) + \mathcal{R} (\mathcal{\tilde{G}}_t, \mathbf{M}_t;\Theta ), 
    \label{eq:image_loss}
\end{split}
\end{equation}
where $\mathcal{R}(x;\phi)$ is the Gaussian 
rasterisation loss with inputs $x$ and optimisable parameters $\phi$, and $\mathcal{\tilde{G}}_t$ are the dynamic Gaussians $\mathcal{G}_t$ with colour $\mathbf{c}$ set to mask foreground (\eg white). %
Note that splatting in the first frame updates shared Gaussian properties, whereas all other frames backpropagate to update the deformation field parameters $\Theta$. 
With the formulation in \cref{eq:surface_point}, our method propagates gradient information from future states to update 3D reconstructions of the earlier frames. 
This enables global space-time surface optimisation while respecting the causality of the surface dynamics. %
While temporal consistency is implicit in \cref{eq:surface_point}, other variants of temporal regularisation are possible (see \cref{sec:temporal_variants}). %
In contrast to previous 3DGS or dynamic Gaussian methods \cite{yang2023deformable}, we note that it is important to optimise all frames in each iteration rather than random frame $t\in[2,...,T]$, as it can otherwise lead to local minima (incorrect learning of folds). 
Global optimisation is preferred as our deformation model is a single global MLP for all frames,
and it is computationally efficient over random frame selection due to the causality of deformations (\cref{eq:surface_point}).
Moreover, we keep the number of Gaussians fixed (\ie,~no adaptive density control) since Gaussian parameters are learned only from the first frame; it is not known beforehand how the deformation details will arise in future states. 
\subsection{Thin Shell Physical Prior} %
\label{ssec:physics_loss}
\input{figures/fig_physics_loss}
If we optimise the NDF solely using the photometric loss, it could perfectly fit the input images but could result in physically implausible surfaces due to monocular ambiguity. 
Therefore, we formulate a physics loss applicable to thin shells inspired by the NeuralClothSim approach~\cite{kairanda2023neuralclothsim} using the Kirchhoff-Love theory~\cite{wempner2003mechanics}. 
Since we model the dynamically tracked surface with an MLP, 
NeuralClothSim enables physics-based prior directly on the continuous surface. 
However, unlike the forward model of NeuralClothSim, the external forces (such as contacts and the wind) and the material properties generating the deformations are unknown in an inverse setting like ours. 
For this reason, we assign material parameters to values typical for the surface modelled (\eg, cloth and paper). 
In contrast to material, it is hard to set reasonable values for forces or boundary conditions as the space of external forces is too large. 
Nevertheless, intuitively, the image loss effectively takes the role of external force as it guides the motion and deformation of the tracked surface.
Thus, we omit the 
potential energy due to external forces from the optimisation. 
Finally, with the strain evaluated from the deformation field and the assumed material, our physical prior aims to minimise the internal hyperelastic energy that captures the surface stretching, shearing and bending stiffness; see \cref{fig:physics_loss}. %

\noindent\textbf{Strain Computation.}
To evaluate stretching and bending strain at each training iteration, 
we randomly sample $N_p$ parametric points $\{\boldsymbol{\xi}_i\}_{i=1}^{N_p}$ from the template mesh $\mathcal{M}$,
and perform differential geometry operations on the initial surface state (\ie,~NRF) and the deformed state (\ie,~NDF). 
We next present the computation of the local quantities at each point $\boldsymbol{\xi}_i$ %
on the initial state using the pre-trained NRF $\mathbf{\bar x}(\boldsymbol{\xi};\Upsilon)$.
We use Greek letters for indexing quantities on the parameterised surface (\textit{e.g.,} $\mathbf{\bar{a}}_{\alpha}, \alpha, \beta, ... = 1,2$). 
For notational clarity, we drop the input $\boldsymbol{\xi}_i,t$ and network weights $\Upsilon,\Theta$ in all the derived quantities (\eg,~$\mathbf{\bar{a}}_1\equiv\mathbf{\bar{a}}_1(\boldsymbol{\xi}_i;\Upsilon), \boldsymbol{\varepsilon}\equiv\boldsymbol{\varepsilon}(\boldsymbol{\xi}_i,t;\Upsilon,\Theta)$). 
In the first step, we extract the local covariant basis: $\{\mathbf{\bar{a}}_\alpha:= \partial \mathbf{\bar x} / \partial \xi^\alpha, \mathbf{\bar{a}}_3:= \mathbf{\bar a}_1 \times \mathbf{\bar a}_2\}$, the set of two vectors tangential to the curvilinear coordinate lines $\xi^\alpha$ and the local normal. 
For the Gaussian samples $\{\boldsymbol{\xi}_i\}_{i=1}^{N_g}$, an identical computation is employed for setting the covariant basis as the rotation parameters (as in \cref{ssec:image_loss}). 
Next, we evaluate the surface metric tensor $\bar{a}_{\alpha \beta}$ measuring the lengths and the curvature tensor $\bar{b}_{\alpha \beta}$ measuring the curvature of the midsurface.
Akin to the covariant tensors (\eg $\bar{b}_{\alpha \beta}$), their contravariant (\eg $\bar{b}^{\alpha \beta}$) and mixed variants (\eg $\bar{b}_{\alpha}^\beta$) counterparts are extracted as well.

With these initial state quantities, 
we next present the strain---due to deformation---as a function of the NDF $\mathbf{u} (\boldsymbol{\xi};\Theta)$. 
The predicted $\mathbf{u}= \hat{u}_i \mathbf{e}_i$ in global Cartesian coordinate system
is transformed to contravariant coordinate basis, $ \mathbf{u} = u_\alpha \mathbf{\bar a}^\alpha + u_3 \mathbf{\bar a}^3 $ for strain calculation \cite{kairanda2023neuralclothsim}. 
Given the deformation gradient $\mathbf{u}_{,\alpha} $ derived from the NDF as 
\vspace{-4pt}
\begin{align}
    \begin{split}
    \mathbf{u}_{,\alpha} &= \varphi_{\alpha \lambda} \mathbf{\bar a}^\lambda + \varphi_{\alpha 3} \mathbf{\bar a}^3, \;\text{with}\; \\
    \varphi_{\alpha \lambda} &:= u_\lambda |_\alpha - \bar{b}_{\alpha \lambda} u_3 \;\text{and}\;
    \varphi_{\alpha 3} := u_{3,\alpha} + \bar{b}_\alpha^\lambda u_\lambda,
    \end{split}
    \label{eq:deformation_gradient}
\end{align}
we evaluate the non-linear membrane strain $\boldsymbol{\varepsilon}= [\varepsilon_{\alpha \beta}]$ and bending strain $\boldsymbol{\kappa}= [\kappa_{\alpha \beta}]$ as
\vspace{-6pt}
\begin{align}
    \small
    \begin{split}
    \varepsilon_{\alpha \beta} &= \frac{1}{2} %
    (\varphi_{\alpha \beta} + \varphi_{\beta \alpha}
    + \varphi_{\alpha \lambda} \varphi_\beta^\lambda + \varphi_{\alpha 3} \varphi_{\beta 3} ), \\
    \kappa_{\alpha \beta} &= 
    -\varphi_{\alpha 3}|_\beta - \bar{b}_\beta^\lambda \varphi_{\alpha \lambda} 
    + \varphi^\lambda_3 (\varphi_{\alpha \lambda} |_\beta + \frac{1}{2} \bar{b}_{\alpha \beta} \varphi_{\lambda 3} - \bar{b}_{\beta \lambda} \varphi_{\alpha 3}), 
    \end{split}
    \label{eq:strain}
\end{align} 
where $\mathbf{\bar a}^i$ denote the local contravariant basis and $\bar{b}_{\alpha \beta}$, the components of the curvature tensor. 
In \cref{eq:strain}, we use a vertical bar for covariant derivatives, lower comma notation for partial derivatives w.r.t.~the curvilinear coordinates $\xi^\alpha$ (\eg $u_\lambda |_\alpha$, and $\mathbf{u,}_\alpha = \partial \mathbf{u} / \partial \xi^\alpha $), and Einstein summation convention of repeated indices for tensorial operations (\eg, $\varphi_{\alpha \lambda} \varphi_\beta^\lambda = \varphi_{\alpha 1} \varphi_\beta^1 + \varphi_{\alpha 2} \varphi_\beta^2$). 
Notably, we compute all the aforementioned physical quantities with automatic differentiation.
Please see \cref{sec:tensor_algebra_suppl} for more details. %

\noindent\textbf{NDF Optimisation.}
As material model, we use $\boldsymbol{\Phi} := \{\rho, h, E, \nu \}$, with mass density $\rho$ and the surface thickness $h$, and elastic coefficients: Young's modulus $E$, and Poisson's ratio $\nu$.
For simplicity and computational efficiency, we use a linear isotropic constitutive model relating strain to stress, thereby %
computing the in-plane stiffness $D$, the bending stiffness $B$ and the elastic tensor 
$\mathbf{H} = [H^{\alpha \beta \lambda \delta}]$ on the initial template, which read as 
\begin{align}
    \begin{split}    
    D &:= \frac{Eh}{1 - \nu^2}\,,\,B := \frac{Eh^3}{12 (1 - \nu^2)}, \;\text{and}\;\\
    H^{\alpha \beta \lambda \delta} &:= \nu \bar{a}^{\alpha \beta} \bar{a}^{\lambda \delta} + \frac{1}{2} (1-\nu) (\bar{a}^{\alpha \lambda} \bar{a}^{\beta \delta} + \bar{a}^{\alpha \delta} \bar{a}^{\beta \lambda}).
    \end{split}
    \label{eq:elastic_tensor}
\end{align}
Finally, our physics loss over the tracked surface states reads as the following: 

\begin{equation}
\small 
\begin{split}
    \mathcal{L}_p(\Theta) &= \frac{1}{2N_p T}\sum_{i=1}^{N_p} \sum_{t=2}^{T} \Big( \underbrace{D \boldsymbol{\varepsilon}^\top (\boldsymbol{\xi}_i, t;\Theta) \mathbf{H} (\boldsymbol{\xi}_i) \boldsymbol{\varepsilon}(\boldsymbol{\xi}_i, t;\Theta)}_{\text{stretching/shearing stiffness}} \\
    & + \underbrace{ B \boldsymbol{\kappa}^\top (\boldsymbol{\xi}_i, t;\Theta) \mathbf{H} (\boldsymbol{\xi}_i) \boldsymbol{\kappa} (\boldsymbol{\xi}_i, t;\Theta) }_{\text{bending stiffness}}\Big)\sqrt{\bar{a}(\boldsymbol{\xi}_i)}, 
\end{split}
\label{eq:physics_loss}
\end{equation}
where $\sqrt{\bar{a}} := |\mathbf{\bar{a}}_1 \times \mathbf{\bar{a}}_2 |$.  
Note that the term inside $\mathcal{L}_p$ evaluates per point; therefore, we re-sample at each training iteration to explore the continuous surface domain.
Finally, we solve for the optimal NDF weights $\Theta^*$ by minimising the objective function $\mathcal{L} = \lambda_{d} \mathcal{L}_{d} + \lambda_{p} \mathcal{L}_{p}$ with empirically determined loss weights $\{\lambda_{d}, \lambda_{p}\}$. 
The optimisation can be interpreted as an inverse simulation, with data loss guiding deformation as an external force and physics loss modelling the surface's intrinsic behaviour. 
We use iterative gradient-based optimisation to that end~\cite{kingma2017adam}. 
We choose $N_p{\ll}N_g$ as the physics loss involves expensive higher-order derivative computations, whereas the Gaussian rasterisation with samples fixed over training iterations is rather efficient. 
With resampling at each iteration, the physics loss backpropagates to the continuous surface, offering adaptive, memory-efficient performance compared to mesh-based physics alternatives requiring high resolution. 

%% file: figures/fig_overview.tex
\begin{figure*}
        \includegraphics[width=\linewidth]{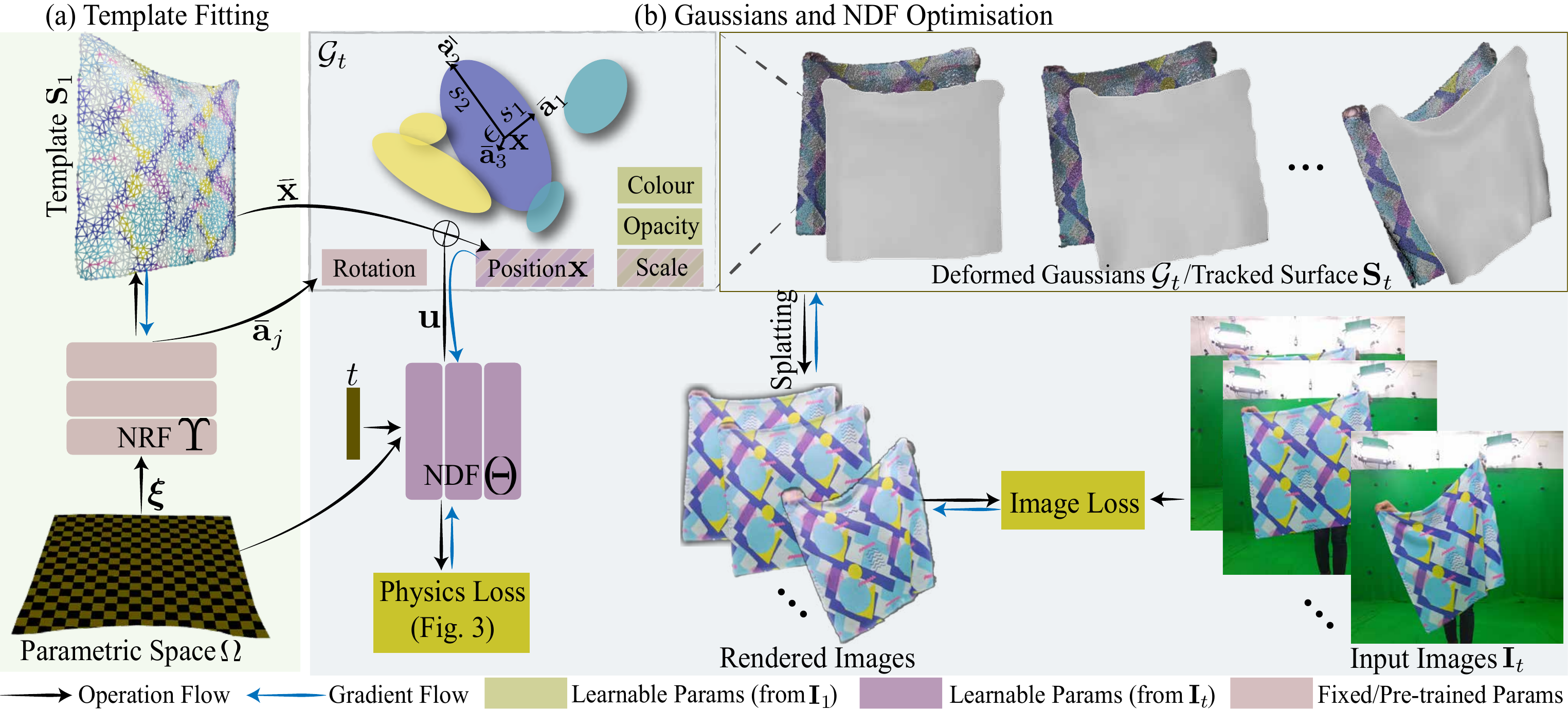}
        \vspace{-16pt}
	\caption
	{
	\textbf{Overview of 
        \ours.} 
      Our deformation model encodes the surface and its dynamics as neural fields. %
      Given the template $\mathbf{S}_1$, we first fit a reference field (NRF) from 2D parametric points $\boldsymbol{\xi}$ to the initial 3D positions $\mathbf{\bar{x}}$. %
      In the main stage, we optimise the %
      deformation field (NDF) $\mathbf{u}(\boldsymbol{\xi},t)$ by relating estimated surface states $\mathbf{S}_t/\mathcal{G}_t$ to the input monocular views. 
      We 
      induce the dynamically tracked Gaussians to the surface by: 
      (1) Computing their positions $\mathbf{x}$ as the sum of the initial position $\mathbf{\bar{x}}$ and NDF output $\mathbf{u}$, (2) Setting their rotations $\mathbf{\bar{a}}_i$ as the template's local coordinate system, and (3) Fixing the normal scale $\epsilon$, and optimising the colour, opacity and tangential scales $(s_1,s_2)$ using only the 
      template texture. 
      For physical plausibility, we impose continuous Kirchhoff-Love physics constraints. 
        }
        \vspace{-14pt}
	\label{fig:overview}
\end{figure*}

%% file: figures/fig_physics_loss.tex
\begin{figure}
        \includegraphics[width=\linewidth]{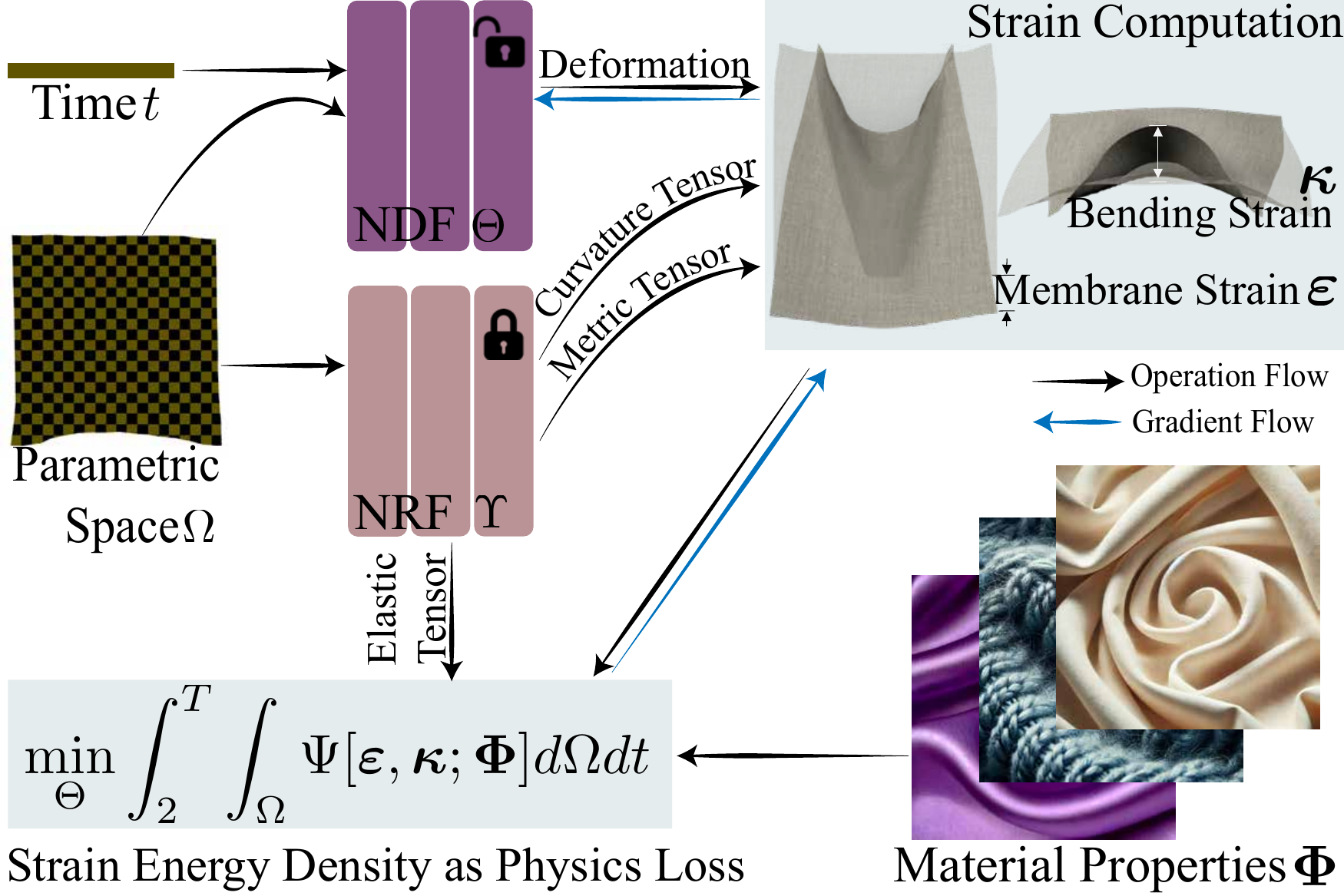} %
	\caption
	{
	\textbf{Thin-shell physics prior} 
        is a spatial regulariser that minimises the hyperelastic strain energy density due to deformation w.r.t.~the known template. 
        }
        \vspace{-18pt}
	\label{fig:physics_loss}
\end{figure}

%% file: sections/4_results.tex
\section{Experimental Results} 
\label{sec:results}

\noindent\textbf{Implementation Details.} 
We implement \ours~using PyTorch~\cite{ravi2020pytorch3d} and optimise separate NRF and NDF networks for each sequence.
The NRF and NDF, both, employ the SIREN \cite{sitzmann2020implicit} architecture using a frequency of $\omega=5$ and $\omega=30$, respectively.
Both networks have five hidden layers and $256$ units in each layer, 
and are optimised using Adam~\cite{kingma2017adam}.
Each network is trained for ${\approx}2 \cdot 10^3$  iterations; first, the template is fitted with NRF, followed by surface tracking with NDF. 
While NRF training takes up to two minutes, the core part of the method, \ie~the NDF training, typically takes between 
30 minutes to one hour until convergence on an NVIDIA A100 GPU. 
The Gaussian scales along the normal are set to $\epsilon=10^{-5}$.
In our experiments, the number of Gaussian samples is $N_g \approx 90k$, %
and the number of samples for physics loss is $N_p = 100$.
Concerning the physics loss, the linear elastic material properties of the surface are set as follows for all sequences: $E=\qty{5000}{Pa}, \nu= 0.25,$ %
and $h=\qty{1.2}{mm}$. 
Moreover, we set the image and physics loss weights to $\lambda_d=5$, and $\lambda_p=1$, whereas the temporal constraint scalar is set to $\lambda=0.4$.
For visualisation of the tracked continuous surface, we generate temporally coherent meshes with ball-pivot meshing~\cite{bernardini1999ball} of the template Gaussian positions $\{\mathbf{\bar x}_i\}_{i\in [1,\ldots,N_g]}$.
We refer to \cref{sec:robustness,,sec:temporal_variants,fig:gaussian_count,fig:smoothness} for studies on the effect of hyperparameters $\omega,\lambda$ and $N_g$.
\input{tables/tab_recon_error_transposed}
\input{figures/fig_normal_comparison}
\input{figures/fig_main_result}

\noindent \textbf{Datasets and Error Metrics.} 
We evaluate on the benchmark \phisft~dataset~\cite{kairanda2022f} consisting of nine RGB videos (and reference depths) of highly challenging cloth deformations. 
As is common with non-rigid reconstruction methods \cite{kairanda2022f, Sidhu2020}, we rigidly align the reconstructions of all compared methods to the ground truth. 
This is achieved with Procrustes alignment \cite{umeyama1991least} for the first frame, which is refined with rigid ICP \cite{besl1992method} for subsequent frames. 
\subsection{Comparison}
\label{subsec:comparisons}
\input{figures/fig_depth_comparison}
\noindent\textbf{SfT and NRSfM.}
We compare \ours~to different SfT methods, namely \phisft~\cite{kairanda2022f}, Yu \etal's Direct, Dense, Deformable (DDD) \cite{Yu2015}, Ngo \etal's Ngo2015 \cite{Ngo2015} and Shimada \etal's IsMo-GAN \cite{Shimada2019}, and finally to Stotko \etal's approach~\cite{stotko2023physics}. 
DDD is provided with the required hierarchy of coarse-to-fine templates, and Ngo2015 and Stotko \etal with the same template as in the \phisft~dataset.
Since the dataset is richly textured, the 2D point tracking methods would perform well; therefore, we additionally compare to the NRSfM methods. 
In particular, we compare to Sidhu \etal's Neural NRSfM (N-NRSfM) \cite{Sidhu2020} and Parashar \etal's Diff-NRSfM \cite{Parashar_2020_CVPR}.
The 2D point correspondences required by the NRSfM methods are provided by densely tracking the 2D points across input images with multi-frame subspace flow (MFSF) \cite{Garg2013MFOF, MFSF}. 
The first frame of the sequence is selected as a keyframe for 2D tracking. 

In \cref{fig:main_result}-(left), we show qualitative reconstructions of \ours~compared to the prior state of the art (SotA) \cite{stotko2023physics, kairanda2022f, Parashar_2020_CVPR}. 
We capture fine wrinkles and folds in the deforming sequences that were not addressed by the earlier works. 
Next, we reconstruct surfaces with severe self-occlusions due to multiple layered folds in the extended 
versions of the \phisft~sequences~\cite{kairanda2022f}. 
See \cref{fig:main_result}-(right), where our physics-based model provides a reasonable prior for occluded regions; 
\phisft~fails here due to prohibitive memory requirements. 
In \cref{tab:numerical_transposed}, we show a quantitative comparison where we report the Chamfer distance (with pseudo-ground-truth point clouds)  against the state-of-the-art methods over all the \phisft~dataset sequences. 
We outperform the existing methods on most sequences, often substantially and on average over all sequences. 
While the default temporal coherency works best on the evaluated dataset, we notice qualitative improvements for two out of the nine sequences with the other variants.
In \cref{fig:normal_comparison}, we further show comparisons between the normal maps of the tracked surface. 
Additional numerical comparisons with the previous SotA~\cite{kairanda2022f} on runtime and normal consistency are in \cref{sec:additional_metrics,tab:runtime_psnr_normal_p2s}. 
We obtain a lower $0.009$ ($\ell_2$) and $0.034$ (cosine) normal error, whereas \phisft~shows $0.013$ and $0.041$, respectively. 
One of the primary reasons for the failure of \phisft~\cite{kairanda2022f} is its limitation to low-resolution meshes 
($\approx$300 vertices) capturing just coarse deformations. 
Our NDF offers, similar to scene representation approaches \cite{mildenhall2020nerf,wang2021neus}, a smooth (\cf~global MLP), low-dimensional (\cf a fixed parameter count) deformation space that is adaptive to dynamic details and can represent high-frequency signals (\cf~sine activation). 
Although we discretise the physics loss, we re-sample at each iteration, leading to an \emph{adaptive} discretisation. 
In contrast, earlier physics-based methods~\cite{kairanda2022f,stotko2023physics} use \emph{fixed} discretisation, either due to differentiable simulation not supporting remeshing~\cite{kairanda2022f} or a fixed-resolution surrogate model~\cite{stotko2023physics}. 
Our continuous formulation enables unprecedented fine-grained results. 

\noindent\textbf{Dynamic View Synthesis.}
Next, we compare with state-of-the-art dynamic view synthesis methods, particularly K-Planes~\cite{fridovich2023k} and Deformable Gaussians~\cite{yang2023deformable}. 
For the latter, we initialise the Gaussians with points sampled from the input template instead of the default SfM points. 
The compared radiance field methods are very effective in novel-view synthesis for monocular videos that provide multi-view cues~\cite{gao2022monocular}. 
However, they fail to recover spatiotemporally coherent surface geometry for monocular RGB videos captured with a single static camera; see \cref{fig:depth_comp}. 

\input{figures/fig_ablation}

\subsection{Ablative Studies}
\label{subsec:ablation}
Next, we test the following aspects of our method: contribution of the physics loss, global 
optimisation and 
surface-induced Gaussians. 
Additional ablations, including momentum and mask loss, are presented in the 
\cref{sec:ablation_suppl}. 

\noindent \textbf{Physics Prior.} 
While relying solely on image loss yields accurate RGB recovery during rendering, it can lead to severely distorted surface tracking. 
This arises from the inherent monocular ambiguity of our setting as multiple possible deformations in 3D correspond to the same 2D image. 
As seen in \cref{fig:ablation}-(a), incorporating Kirchhoff--Love-based thin shell prior is essential for achieving physically plausible and accurate surface reconstruction. 

\noindent \textbf{Global Optimisation.} %
We test a version of our method employing optimisation with randomly selected frames at each iteration 
instead of the proposed joint optimisation over all frames. 
\cref{fig:ablation}-(b) %
visualises the ablated results showing poor performance for folds and wrinkles. 
Without any multi-view cues, as in the case of the used  \phisft~dataset, random frame optimisation leads to local minima. 

\noindent \textbf{Surface-induced Gaussians.} 
The specially-tailored initialisation and optimisation of Gaussian parameters (\cref{eq:tracked_Gaussian,eq:image_loss}) are crucial 
for the accurate geometric reconstruction. 
In the ablated version of our method, we optimise the shared Gaussian parameters, \ie, covariance, opacity, and colour on all input frames (similar to earlier methods~\cite{kerbl20233d,yang2023deformable}) instead of the single template frame as in \cref{eq:image_loss}.
Due to the ambiguities between geometric details and appearance over the deforming sequence, this leads to wrongly reconstructed Gaussians, 
eventually resulting in poor surface reconstructions; see \cref{fig:ablation}-(c). 
In addition, fixing the normal scale on the template frame is useful for preventing elongated Gaussians (\cref{fig:ablation}-(d)).

In \cref{tab:ablation_transposed}-appendix, we report the above results on the full \phisft~dataset. %
On average, we obtain Chamfer distances of $34.25$, $14.0$, $3.75$, and $3.46$ for no physics, 
no surface-induced Gaussians, no normal scale, and the full model, respectively. 
We notice that including physics and surface induction are crucial for accurate tracking. %

%% file: tables/tab_recon_error_transposed.tex
\begin{table}[tb]  
  \centering
  \begin{adjustbox}{width=\linewidth,center}
  \begin{tabular}{rrrrrrrrrrr}
   \toprule
    Seq.    & S1    & S2    & S3    & S4    & S5    & S6    & S7    & S8    & S9     & \textbf{Avg} \\
    \midrule
    N-NRSfM~\cite{Sidhu2020}                     & 8.25 & 33.62 & 104.6 & 77.02 & 72.66 & 8.73 & 129.4 & 38.06 & 19.81 & \textbf{54.69} \\
    D-NRSfM~\cite{Parashar_2020_CVPR}            & 17.14 & 4.46 & 4.40 & 41.37 & 26.92 & 14.02 & 12.49 & 9.91 & \cellcolor{third} 5.29 & \textbf{15.11} \\
    \midrule
    Shimada~\cite{Shimada2019}                  & 19.69 & 22.18 & 33.54 & 90.30 & 92.78 & 57.62 & 49.27 & 24.45 & 53.12 & \textbf{49.22} \\
    DDD~\cite{Yu2015}                            & 2.95 & 1.69 & 3.80 & 25.73 & \cellcolor{third} 10.46 & 6.97 & 15.64 & \cellcolor{third} 7.61 & 11.77 & \textbf{10.87} \\
    Ngo ~\cite{Ngo2015}                          & \cellcolor{third} 2.19 & \cellcolor{second} 1.51 & \cellcolor{first} 2.17 & \cellcolor{third} 15.90 & 10.72 & \cellcolor{second} 3.01 & 7.95* & fail & fail & \cellcolor{third} \textbf{5.92*} \\    
    Stotko~\cite{stotko2023physics}              & 6.1 & 3.9 & 12.5 & 14.5 & 11.7 & 15.1 & \cellcolor{third} 6.9 & 10.1 & 8.6 & \textbf{9.93}\\
    $\boldsymbol{\phi}$-SfT~\cite{kairanda2022f} & \cellcolor{first} 0.79 & \cellcolor{third} 2.75 & \cellcolor{third} 3.54 & \cellcolor{second} 7.60 & \cellcolor{first} 6.15 & \cellcolor{third} 3.14 & \cellcolor{second} 4.73 & \cellcolor{second} 2.52 & \cellcolor{first} 2.36 & \cellcolor{second} \textbf{3.93}\\
    \midrule
      Ours                                        & \cellcolor{second} 1.17 & \cellcolor{first} 0.55 & \cellcolor{first} 2.4\dag/3.5 & \cellcolor{first} 5.5\dag/5.7 & \cellcolor{second} 8.69 & \cellcolor{first} 2.51 & \cellcolor{first} 3.8 & \cellcolor{first} 2.27 & \cellcolor{second} 3.00 & \cellcolor{first} \textbf{3.3\dag/3.5}\\      
    \bottomrule
  \end{tabular}
  \end{adjustbox}
 \vspace{-8pt}
  \caption{%
  We quantitatively compare \ours~to the state of the art on the $\boldsymbol{\phi}$-SfT real dataset. 
  The average Chamfer distance is multiplied by $10^4$ for readability. %
  ``$^*$'' notes that Ngo \etal failed on the last few frames of S7, which we exclude from the error computation.
  ``$\dag$'' denotes that we report the numbers on the variant of temporal coherency constraint described in \cref{sec:temporal_variants}. 
  }
  \label{tab:numerical_transposed}
   \vspace{-10pt}
\end{table}

%% file: figures/fig_normal_comparison.tex
\begin{figure}
  \centering
  \includegraphics[width=\linewidth]{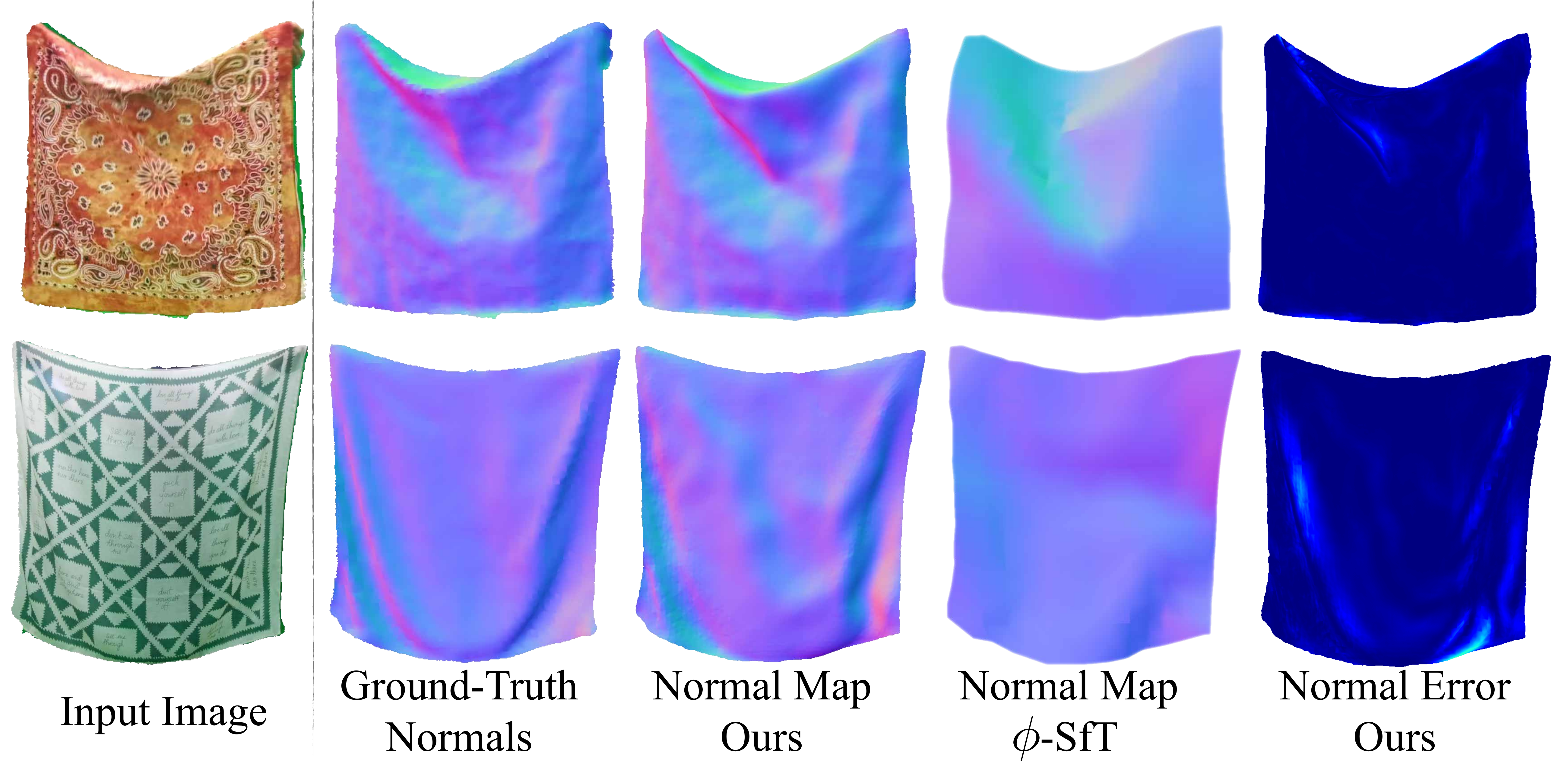}
  \vspace{-20pt}
   \caption{ 
    \textbf{Comparison of the reconstruction normal maps} for ours and \phisft. 
   We show the cosine normal consistency to ground truth on the right and the normal metrics in \cref{tab:runtime_psnr_normal_p2s} (Appendix). 
   }
   \label{fig:normal_comparison}
   \vspace{-18pt}
\end{figure}

%% file: figures/fig_main_result.tex
\begin{figure*}
        \includegraphics[width=1\linewidth]{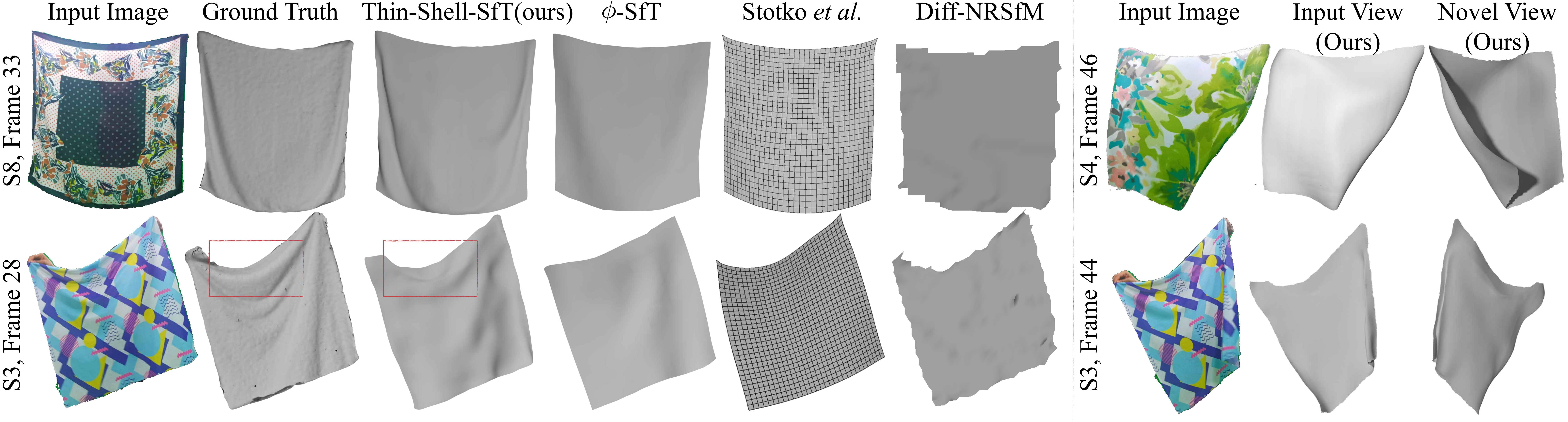} %
        \vspace{-22pt}
	\caption
	{
	\textbf{Examplary 3D reconstructions.} 
        (Left:) Comparisons focusing on high-frequency wrinkles. 
        \ours~captures the wrinkles best among all compared methods in one of the most challenging examples, outperforming $\phi$-SfT \cite{kairanda2022f}, Stotko \etal \cite{stotko2023physics} and Diff-NRSfM~\cite{Parashar_2020_CVPR}. 
        (Right:) Our results on the extended \phisft~dataset highlight the excellent tracking in the occluded regions. 
        }
        \vspace{-18pt}
	\label{fig:main_result}
\end{figure*}

%% file: figures/fig_depth_comparison.tex
\begin{figure}
        \includegraphics[width=\linewidth]{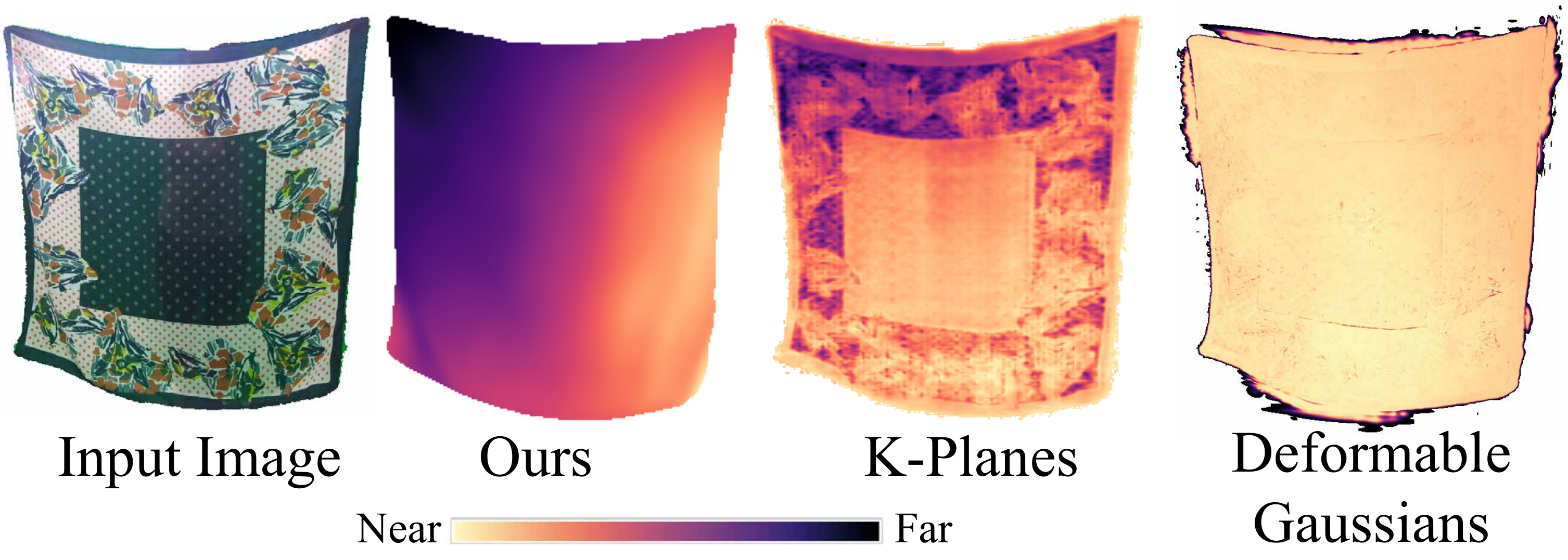} %
        \vspace{-18pt}
	\caption %
        {
        We compare our reconstructions to dynamic view synthesis methods \cite{fridovich2023k, yang2023deformable} on the rendered depth maps. The texture details retained in the depth maps imply that the compared methods fail to learn accurate surface (mistake texture for geometry).}
        \vspace{-10pt}
	\label{fig:depth_comp}
\end{figure}

%% file: figures/fig_ablation.tex
\begin{figure}
        \includegraphics[width=\linewidth]{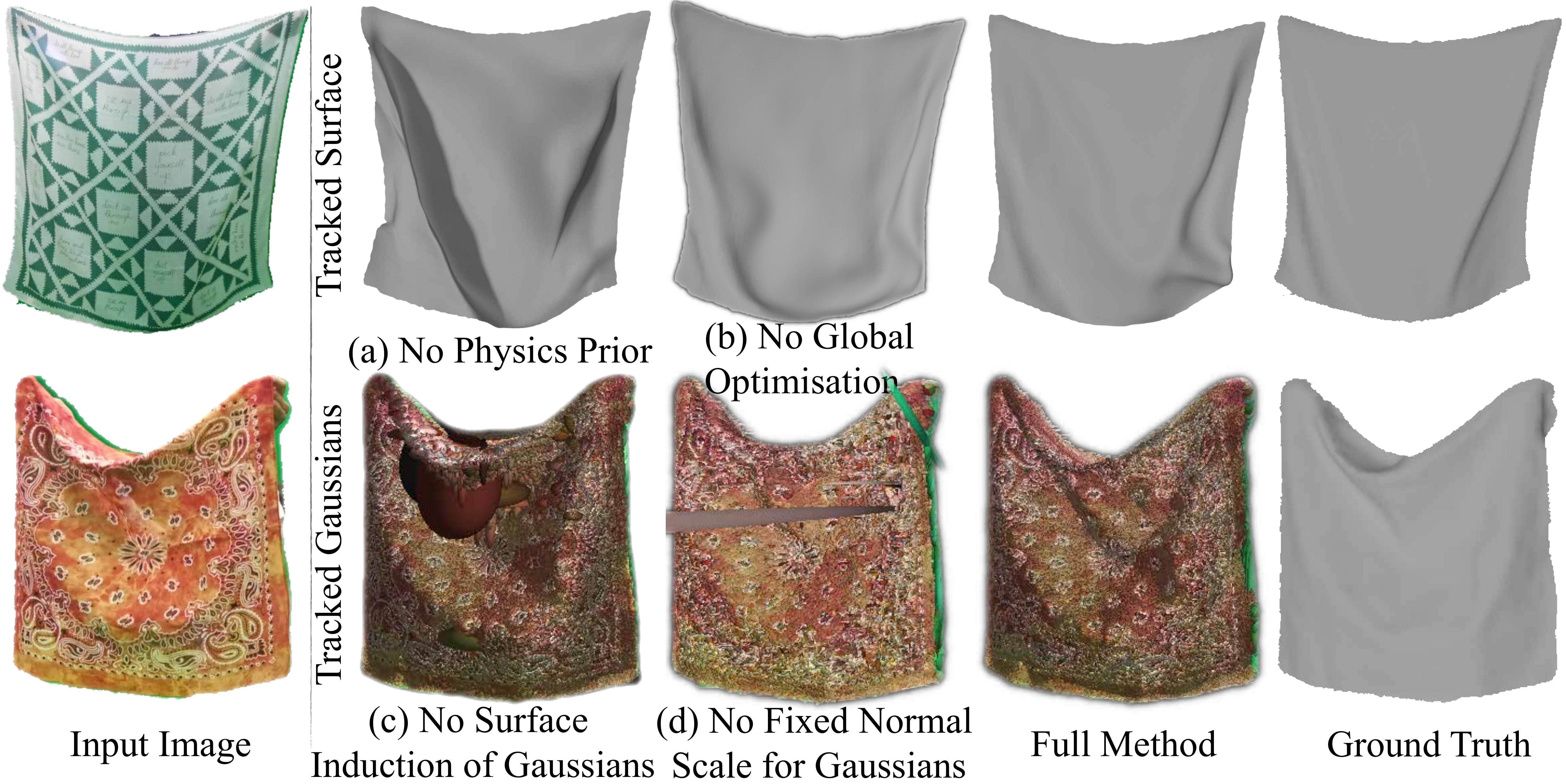} %
        \vspace{-16pt}
	\caption
        {
        \textbf{Ablations.} %
        (a:) Without physics, the surface can severely stretch or shrink. %
        (b:) Random frame optimisation leads to local minima,
        (c,d:) 
        Not positioning and orienting Gaussians on the surface leads to distortions and poor reconstructions. %
        }
        \vspace{-22pt}
	\label{fig:ablation}
\end{figure}

%% file: sections/5_conclusion.tex
\section{Conclusion}
\label{sec:conclusion} 
We present \ours, a new approach for dense monocular non-rigid 3D surface tracking relying on new principles, \ie~\emph{adaptive} neural deformation field, \emph{continuous} Kirchhoff-Love thin shell prior and \emph{surface-induced} 3D Gaussian Splatting. 
Experiments on the \phisft~dataset demonstrate that our method accurately reconstructs the challenging \emph{fine-grained} deformations such as cloth folds and wrinkles. 
Overall, we significantly improve fine-grained surface tracking using an adaptive deformation model and continuous thin shell physics compared to existing approaches that only support coarse deformations. 
Thanks to the physics prior, \ours~is reasonably robust to occlusions, although extreme self-collisions remain challenging. 
Similarly, tracking textureless surfaces is another possible research direction that would necessitate special handling in future. 

%% file: sections/7_acknowledgements.tex
\\
\noindent{\textbf{Acknowledgement.}
This work was partially supported by the VIA Research Center. 
}

%% file: sections/6_appendix.tex
\clearpage
\maketitlesupplementary
\appendix
\setcounter{figure}{0}
\setcounter{table}{0}
\renewcommand*\thetable{\Roman{table}}
\renewcommand*\thefigure{\Roman{figure}}
\renewcommand{\contentsname}{Table of Contents}
\startcontents[sections]
\printcontents[sections]{ }{1}{\section*{\contentsname}\setlength{\cftsecnumwidth}{1.75em}}
\vspace{5mm} 

\section{Thin Shell Physical Prior}
\label{sec:tensor_algebra_suppl}
This section provides additional details on the physics prior, focusing on the differential geometric computations on the parameterised initial and deformed surfaces. 
First, we write down the detailed notations used in the main matter and this appendix. 
Next, we describe the geometric quantities and computations required for evaluating strain in \cref{eq:strain}-(main matter), and subsequently the physics loss (\cref{eq:physics_loss}-(main matter)). 
\paragraph{Notations.} 
Following NeuralClothSim~\cite{kairanda2023neuralclothsim}, we use Greek letters for indexing quantities on the 2D-dimensional surface (\textit{e.g.,} $\mathbf{a}_{\alpha}, \alpha, \beta, ... = 1,2$).
An index can appear as a superscript or subscript. 
Superscripts $(\cdot)^\alpha$ refer to contravariant tensor components, which scale inversely with the change of basis; subscripts $(\cdot)_\alpha$ refer to covariant components that change in the same way as the basis scale. 
We use upper dot notation for time derivatives; vertical bar for covariant derivatives; and lower comma notation for partial derivatives w.r.t.~the curvilinear coordinates, $\xi^\alpha$ (\textit{e.g.,} %
$\dot{\mathbf{u}} = \partial \mathbf{u} / \partial t, u_\lambda |_\alpha$, and $\mathbf{u,}_\alpha = \partial \mathbf{u} / \partial \xi^\alpha $, respectively). 
Moreover, geometric quantities with overbar notation $\bar{(\cdot)}$ refer to the initial surface state, and Einstein summation convention of repeated indices is used for tensorial operations (\textit{e.g.,} $\varphi_{\alpha \lambda} \varphi_\beta^\lambda = \varphi_{\alpha 1} \varphi_\beta^1 + \varphi_{\alpha 2} \varphi_\beta^2$). 
For notational clarity, we drop the input $\boldsymbol{\xi},t$ and parameters $\Upsilon,\Theta$ in all the derived quantities (\eg,~$\mathbf{\bar{a}_1}(\boldsymbol{\xi};\Upsilon), \boldsymbol{\varepsilon}(\boldsymbol{\xi},t;\Upsilon,\Theta)$).
\paragraph{\textbf{Covariant Basis.}}
In the first step, we define a local covariant basis to express local quantities such as the metric and curvature tensors on the initial surface  $\mathbf{\bar x}$. 
This basis includes $\mathbf{\bar a}_\alpha$, the set of two vectors tangential to the curvilinear coordinate lines $\xi^\alpha$: 
\begin{align}\label{eq:covariant_basis} 
    \mathbf{\bar a}_\alpha := \mathbf{\bar x,}_\alpha. 
\end{align} 
The local unit normal $\mathbf{\bar a}_3$, is then computed as the cross product of the tangent base vectors: 
\begin{align}
\mathbf{\bar a}_3 := \frac{\mathbf{\bar a}_1 \times \mathbf{\bar a}_2}{|\mathbf{\bar a}_1 \times \mathbf{\bar a}_2 |},\;\;\mathbf{\bar a}^3 = \mathbf{\bar a}_3.  
\end{align} 
The local basis $\{\mathbf{\bar a}_1,\mathbf{\bar a}_2,\mathbf{\bar a}_3\}$ is additionally used as per-point rotation matrix $\mathbf{R}$ for the Gaussian tracking (see \cref{ssec:image_loss}).
The surface area differential $\,d\Omega$ relates to the curvilinear coordinates via the Jacobian of the metric tensor: 
\begin{align}
    \,d\Omega = \sqrt{\bar a} \,d \xi^1 \,d \xi^2,\;\text{where}\;\sqrt{\bar a} := |\mathbf{\bar a}_1 \times \mathbf{\bar a}_2 |.
\end{align}
\paragraph{\textbf{Metric Tensor and Contravariant Basis.}}
The covariant components of the symmetric metric tensor (\ie, first fundamental form) that measures
the distortion of length and angles are computed as: 
\begin{align}
\bar a_{\alpha \beta} = \bar a_{\beta \alpha} := \mathbf{\bar a}_\alpha \cdot \mathbf{\bar a}_\beta. 
\end{align}
The corresponding contravariant components of the symmetric metric tensor denoted by $\bar a^{\alpha \lambda}$ are obtained using the identity: $\bar a^{\alpha \lambda} \bar a_{\lambda \beta} = \delta_{\alpha \beta}$,  
where $\delta_{\alpha \beta}$ stands for the Kronecker delta. 
$\bar a^{\alpha \lambda}$ can be used to compute the contravariant basis vectors
as follows: $\mathbf{\bar a}^\alpha = \bar a^{\alpha \lambda} \mathbf{\bar a}_\lambda$.
While the covariant base vector $\mathbf{\bar a}_\alpha$ is tangent to the $\xi^\alpha$ line, the contravariant base vector $\mathbf{\bar a}^\alpha$ is normal to $\mathbf{\bar a}_\beta$ when $\alpha \neq \beta$. 
Note that $\mathbf{\bar a}_\alpha$ and $\mathbf{\bar a}^\alpha$  are not necessarily unit vectors.
\paragraph{\textbf{Curvature Tensor.}}
The curvature metric of the initial surface (\ie the second fundamental form) is computed as follows: 
\begin{align}\label{eq:second_ff} 
\bar b_{\alpha \beta} &:= -\mathbf{\bar a}_\alpha \cdot \mathbf{\bar a}_{3,\beta} = -\mathbf{\bar a}_\beta \cdot \mathbf{\bar a}_{3,\alpha} =  \mathbf{\bar a}_{\alpha, \beta} \cdot \mathbf{\bar a}_3. 
\end{align}

\paragraph{\textbf{Covariant Derivatives.}} %
When taking derivatives along a curve on the midsurface, we must account for the change of the local basis along that curve. 
More concretely, we rely on the \emph{surface covariant derivative} to evaluate the deformation gradient $\mathbf{u},_\alpha$ on the deformed midsurface in~\cref{eq:deformation_gradient} of the main paper. %
We compute the covariant derivatives of the \emph{deformed surface} quantities, \ie first-order tensor $u_\lambda |_\alpha$ and the second-order tensor $\varphi_{\alpha \lambda} |_\beta$ in \cref{eq:deformation_gradient,eq:strain}-(main matter), 
using the following rules: 
\begin{align}
\begin{split}
& u_\alpha |_\beta = u_{\alpha,\beta}  - u_\lambda \Gamma_{\alpha \beta}^\lambda, \text{ and } \\
& \varphi_{\alpha \beta}|_\gamma = \varphi_{\alpha \beta, \gamma} - \varphi_{\lambda \beta} \Gamma_{\alpha \gamma}^\lambda - \varphi_{\alpha \lambda } \Gamma_{\beta \gamma}^\lambda,  
\end{split}
\end{align}
where the Christoffel symbol $\Gamma_{\alpha \beta}^\lambda$ is defined as (similarly for $\Gamma_{\alpha \gamma}^\lambda$ and $\Gamma_{\beta \gamma}^\lambda$),
\begin{align}
\Gamma_{\alpha \beta}^\lambda := \mathbf{\bar{a}}^\lambda \cdot \mathbf{\bar{a}}_{\alpha, \beta}. 
\end{align}
\paragraph{\textbf{Symmetric Tensors.}} %
We exploit the symmetry with respect to indices $\alpha$ and $\beta$, \ie $a_{\alpha \beta} = a_{\beta \alpha }$, for efficient computations of the following tensors: $\bar a_{\alpha \beta}$, $\bar b_{\alpha \beta}$, $\varepsilon_{\alpha \beta}$,  $\kappa_{\alpha \beta}$, and $\Gamma_{\alpha \beta}^\lambda$. The fourth-order symmetric tensor $\mathbf{H}$, as in \cref{eq:elastic_tensor}-(main matter), uses:
\[
    H^{\alpha \beta \lambda \delta} = H^{\beta \alpha \lambda \delta} = H^{\beta \alpha \delta \lambda} = H^{\alpha \beta \delta \lambda} = H^{\lambda \delta \alpha \beta}. 
\]
This property means that only six independent components (after applying symmetry) need to be computed (\textit{i.e.,} $H^{1111}$, $H^{1112}$, $H^{1122}$, $H^{1212}$, $H^{1222}$, and $H^{2222}$). 
\input{figures/fig_psnr}
\input{tables/tab_runtime_psnr_normal_p2s}
\section{Variants of the Temporal Constraint}
\label{sec:temporal_variants}
We proposed a momentum regulariser in our deformation formulation (\cref{eq:surface_point}-(main matter)). %
Along with this, as mentioned in the main matter, we experimented with two other variants of temporal consistency that gave improved qualitative results for two of the nine \phisft~sequences (S3 and S4). 
For these variants, we reformulated  the deformed point position on the tracked surface as 
\begin{equation}
\begin{split}
        \mathbf{x}(\boldsymbol{\xi}, t) &= \mathbf{\bar{x}}(\boldsymbol{\xi}) + \mathbf{u}(\boldsymbol{\xi}, t) \\
        \text{ with }\; \mathbf{u}(\boldsymbol{\xi}, t) &= \mathcal{F}(\boldsymbol{\xi}, t), \forall t \in [1,...,T], 
\end{split}
\end{equation}
where we directly regress the deformation (NDF) as the offset to the initial state using MLP $\mathcal{F}(\cdot)$.
As $\mathbf{u}(\boldsymbol{\xi}, 1)=0$ is no longer implicit (unlike \cref{eq:surface_point}), the total loss $\mathcal{L}$ now additionally includes minimisation objectives of (a) initial deformation $\mathbf{u}(\boldsymbol{\xi}, 1;\Theta)$ and (b) either acceleration $\ddot{\mathbf{u}}(\boldsymbol{\xi}, t;\Theta)$ (variant I, S3) or velocity $\dot{\mathbf{u}}(\boldsymbol{\xi}, t;\Theta)$ (variant II, S4).

\noindent\textbf{Regarding $\boldsymbol{\lambda}$}. For the momentum regulariser (\cref{eq:surface_point}-(main matter)), we tried $\lambda{=}1$ instead of the proposed value $\lambda{=}0.4$ in our early experiments. 
In that case, the network prediction $F(\boldsymbol{\xi},t)$ would have an alternate interpretation of velocity instead of deformation offset. 
However, this led to noisier initialisation of the later surface states
due to accumulated offset and, hence, noisy optimisation; thus, we decided upon a $\lambda{<}1$.
Note that $\lambda$ is positive to encourage deformation follow-through (more details in \cref{sec:ablation_suppl}).

\section{Additional Evaluations}
\label{sec:additional_metrics}
\noindent\textbf{Dynamic Novel View Synthesis.}
Although our work focuses on deformable surface tracking but not directly novel view synthesis or appearance reconstruction, 
we additionally show textured tracking and compute the PSNR and LPIPS from input views (ground truth is not available for novel views); see \cref{fig:psnr,tab:runtime_psnr_normal_p2s}. 

\noindent\textbf{Normal Maps.}
In addition to the Chamfer distance (\cref{tab:numerical_transposed}), we evaluate our reconstructions with another image-based metric, \ie cosine and $\ell_2$ normal consistency (following Refs.~\cite{he2020geo,sun2024metacap}); see \cref{tab:runtime_psnr_normal_p2s} and \cref{fig:normal_comparison}-(main matter) for all results.
The normal metric captures the error in the fine-grained details of the reconstructions, where we notably outperform the previous SotA (\phisft).

\noindent\textbf{Runtime.}
In~\cref{tab:runtime_psnr_normal_p2s}, we report the runtime for each sequence.
\ours~typically takes between 30 minutes and one hour until convergence on an NVIDIA A100 GPU.
Although computationally expensive, ours is significantly faster ($\approx38\times$) than \phisft~\cite{kairanda2022f}, which takes up to $16{-}24$ hours. 
While a recent method~\cite{stotko2023physics} takes up to three minutes, 
our method significantly outperforms both in the fine-grained wrinkle reconstruction.
\input{figures/fig_smoothness}
\input{figures/fig_gaussian_count}
\section{Hyperparameters} 
\label{sec:robustness}
\noindent \textbf{Number of Gaussians.}
In \cref{fig:gaussian_count}, we report the reconstruction error, visualise the surfaces for varying Gaussian counts, and observe the reconstruction quality drops only slightly with fewer  Gaussians. %
The number of Gaussian samples in our main experiments is $N_g \approx 80{-}90k$. 

\noindent \textbf{Smoothness Control.}
Depending on the application, it could be useful to control the smoothness and sharpness of the reconstructed surface. %
This can be achieved by tuning the frequency $\omega$ (set to default 30 in all our experiments) of sinusoidal activation~\cite{sitzmann2020implicit} in the NDF; see \cref{fig:smoothness}. %

\section{Detailed Ablations}
\label{sec:ablation_suppl}
\input{tables/tab_ablation_transposed}
We provided a detailed description (\cref{subsec:ablation}), the qualitative results (\cref{fig:ablation}) and summarised ablation results in the main matter. Here, we additionally report the results on the full \phisft~dataset and provide three additional ablations. 
We test the following modes: 1) Without Kirchhoff-Love thin-shell-based physical prior, 2) No surface-induced Gaussians, \ie, optimising Gaussian parameters (\ie, scale, opacity, and colour) on all input frames instead of the single (template) frame, and 3) Without fixing the scale along the surface normal, 4) Without the momentum regularisation, and 5) Without mask loss.
In \cref{tab:ablation_transposed}, we report the error to the ground truth for all the ablated versions on each sequence.
We notice that including \emph{continuous} physics loss and surface-induced Gaussians are crucial for accurate surface tracking. %

\noindent \textbf{No Fixed Normal Scale.} 
Regarding the surface-induced Gaussians, %
we test the variant that optimises the 3D scales $ (s_1, s_2, s_3)_i$ and rotation $\mathbf{R}_i$ of each Gaussian in $\mathcal{G}_1$ instead of setting $s_3 :=\epsilon$ and $\mathbf{R}_i=[\mathbf{\bar a}_1 \; \mathbf{\bar a}_2 \; \mathbf{\bar a}_3]_i$, as in our full method. 
1) Missing normal scale regularisation leads to elongated 3D Gaussians along the view direction, leading to a high RGB loss; 
see \cref{fig:ablation}-(d)-(main matter).

\noindent \textbf{Momentum Regularisation.} 
We perform joint space-time NDF optimisation while enforcing backpropagation of information to previous states using (\cref{eq:surface_point}-(main matter)). 
By setting $\lambda = 0$, we test the variant %
with no explicit temporal constraint. 
This reduces the accuracy as reported in \cref{tab:ablation_transposed}. %
The momentum term encourages the current deformed state to follow the previous deformation. 
It especially helps in sequences with large 
sway 
(\eg,~single-wrinkled S2) and is less effective for 
frequently alternating deformations. 

\noindent \textbf{Mask Loss.}
Masks are optional inputs for our method. %
When using mask loss, we observe a speedup in convergence ($1.5\times$ faster) but did not notice much qualitative or quantitative improvement in surface tracking. 

\input{figures/fig_additional_results}
\section{Qualitative Results} 
\label{sec:suppl_qualitative}
Our \ours~outperforms the existing methods, especially qualitatively and for \emph{fine-grained} details such as wrinkles.
We visualise reconstructions of our \ours~on two \phisft~sequences; see \cref{fig:suppl_result}. 
The figure shows the input image sequences of the evolving surface and their corresponding spatiotemporally coherent 3D reconstructions for selected frames.
Please refer to the supplemental video for the visualisation of surface tracking of all sequences. %
\section{Limitations}
\label{sec:limitations}
\input{figures/fig_limitation}
Our method reconstructs the challenging fine-grained surface deformations from monocular videos.
Thanks to the physics prior, the method is reasonably robust to occlusions although we notice self-collision in extreme cases, as this is not explicitly handled; see \cref{fig:limitation}. 
Since the surface can cast self-shadows, non-Lambertian surfaces can appear differently over time.
While our approach remains robust against changes in appearance across frames for the tested dataset, substantial changes (\eg~specular surfaces) can lead to a decline in 3D reconstruction quality. 
Similarly, tracking of textureless surfaces is yet another important problem; we leave it as future work.
Overall, we significantly improve surface tracking using an \emph{adaptive} deformation model, \emph{continuous} thin shell loss and surface-induced 3D Gaussian Splatting compared to existing approaches.

%% file: figures/fig_psnr.tex
\begin{figure}
  \centering
  \includegraphics[width=\linewidth]{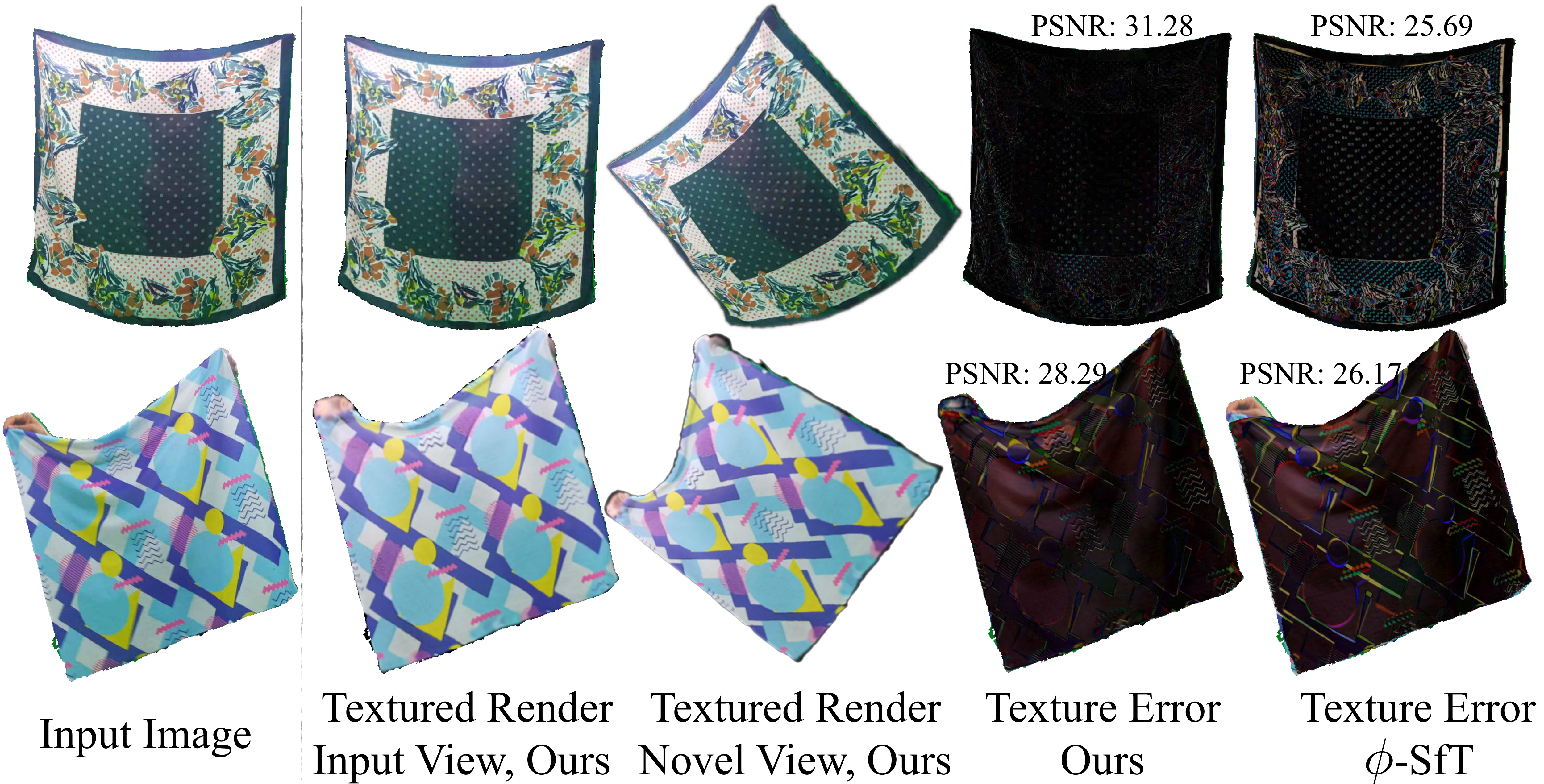}
   \caption{
   \textbf{Dynamic novel-view synthesis.} We render the tracked Gaussians from input and novel views
   and visualise the texture error ($\ell_1$ loss) for the input view. %
   Our lower texture error compared to previous SotA enables higher-fidelity surface reconstructions.
   }
   \label{fig:psnr}
\end{figure}

%% file: tables/tab_runtime_psnr_normal_p2s.tex
\begin{table}[tb]  
  \centering
  \begin{adjustbox}{width=\linewidth,center}  
  \begin{tabular}{llrrrrrrrrrrr}
   \toprule
    Metric    & Method &  S1    & S2    & S3    & S4    & S5    & S6    & S7    & S8    & S9     & \textbf{Avg} \\
    
    \midrule
    \multirow{2}{3.9em}{NC-Cos$\downarrow$}  
     & \phisft    &  \cellcolor{first} 0.022 & \cellcolor{second} 0.064 & \cellcolor{first} 0.059 & \cellcolor{second} 0.037 & \cellcolor{first} 0.044  & \cellcolor{second} 0.028 & \cellcolor{second} 0.058 & \cellcolor{second} 0.030 &  \cellcolor{second} 0.026 & \cellcolor{second} \textbf{0.041}\\
     & Ours       &  \cellcolor{second}0.029 &  \cellcolor{first} 0.015 & \cellcolor{second}0.071 & \cellcolor{first} 0.036 & \cellcolor{second}0.062 & \cellcolor{first} 0.014 & \cellcolor{first} 0.039 & \cellcolor{first} 0.017 & \cellcolor{first}0.022 & \cellcolor{first}\textbf{0.034}\\
    
    \midrule
    \multirow{2}{3.5em}{NC-$\ell_2\downarrow$}  
     & \phisft    &  \cellcolor{second}0.006 & \cellcolor{second}0.011 & \cellcolor{first}0.014 & \cellcolor{second}0.013 & \cellcolor{second}0.016 & \cellcolor{second}0.013 & \cellcolor{second}0.016 & \cellcolor{second}0.013 & \cellcolor{second}0.012 & \cellcolor{second}\textbf{0.013}\\
    &  Ours  & \cellcolor{first}0.005 & \cellcolor{first}0.006 & \cellcolor{second}0.016 & \cellcolor{first}0.009 & \cellcolor{first}0.015 & \cellcolor{first}0.007 & \cellcolor{first}0.010 & \cellcolor{first}0.008 & \cellcolor{first}0.008 & \cellcolor{first}\textbf{0.009}\\
    
    \midrule
    \multirow{2}{3.5em}{PSNR$\uparrow$} 
     & \phisft    & \cellcolor{second}28.17 & \cellcolor{second}26.82 & \cellcolor{second}26.17 & \cellcolor{second}27.49 & \cellcolor{second}25.18 & \cellcolor{second}25.08 & \cellcolor{second}23.15 & \cellcolor{second}25.69 & \cellcolor{second}26.39 & \cellcolor{second}\textbf{26.02}\\
     &     Ours   & \cellcolor{first}32.83 & \cellcolor{first}31.93 & \cellcolor{first}28.29 & \cellcolor{first}30.81 & \cellcolor{first}30.22 & \cellcolor{first}31.24 & \cellcolor{first}29.09 & \cellcolor{first}31.28 & \cellcolor{first}31.34 & \cellcolor{first}\textbf{30.78}\\
     
     \midrule
    \multirow{2}{3.5em}{LPIPS$\downarrow$} 
     & \phisft    & \cellcolor{second}0.021 & \cellcolor{second}0.027 & \cellcolor{first}0.040 & \cellcolor{first}0.045 & \cellcolor{second}0.055 & \cellcolor{first}0.048 & \cellcolor{first}0.049 & \cellcolor{second}0.049 & \cellcolor{second}0.045 & \cellcolor{first}\textbf{0.042}\\
     &     Ours   & \cellcolor{first}0.013 & \cellcolor{first}0.021 & \cellcolor{second}0.052 & \cellcolor{second}0.049 & \cellcolor{first}0.043 & \cellcolor{second}0.066 & \cellcolor{second}0.054 & \cellcolor{first}0.042 & \cellcolor{first}0.043 & \cellcolor{first}\textbf{0.042}\\
     
    \midrule
    \multirow{2}{3.6em}{Runtime$\downarrow$}  & \phisft &  \cellcolor{second}20h3m & \cellcolor{second}6h23m & \cellcolor{second} 15h45m & \cellcolor{second}21h33m  & \cellcolor{second} 9h10m & \cellcolor{second} 5h25m & \cellcolor{second} 9h40m  & \cellcolor{second} 17h1m  & \cellcolor{second}18h3m   & \cellcolor{second} \textbf{13h40m}\\ 
     &   Ours  & \cellcolor{first}47m & \cellcolor{first}28m & \cellcolor{first}38m & \cellcolor{first}36m & \cellcolor{first}43m & \cellcolor{first}34m & \cellcolor{first}36m & \cellcolor{first}37m & \cellcolor{first}39m & \cellcolor{first}\textbf{38m} \\
    
    \bottomrule
  \end{tabular}
  \end{adjustbox}
  \caption{%
  \textbf{Additional metrics and comparisons.} We compare with \phisft~on image-based metrics, including cosine and $\ell_2$ normal consistency error; %
   our \ours~generates more accurate results while being significantly faster than \phisft.}
  \label{tab:runtime_psnr_normal_p2s}
\end{table}

%% file: figures/fig_smoothness.tex
\begin{figure}
  \centering
  \includegraphics[width=\linewidth]{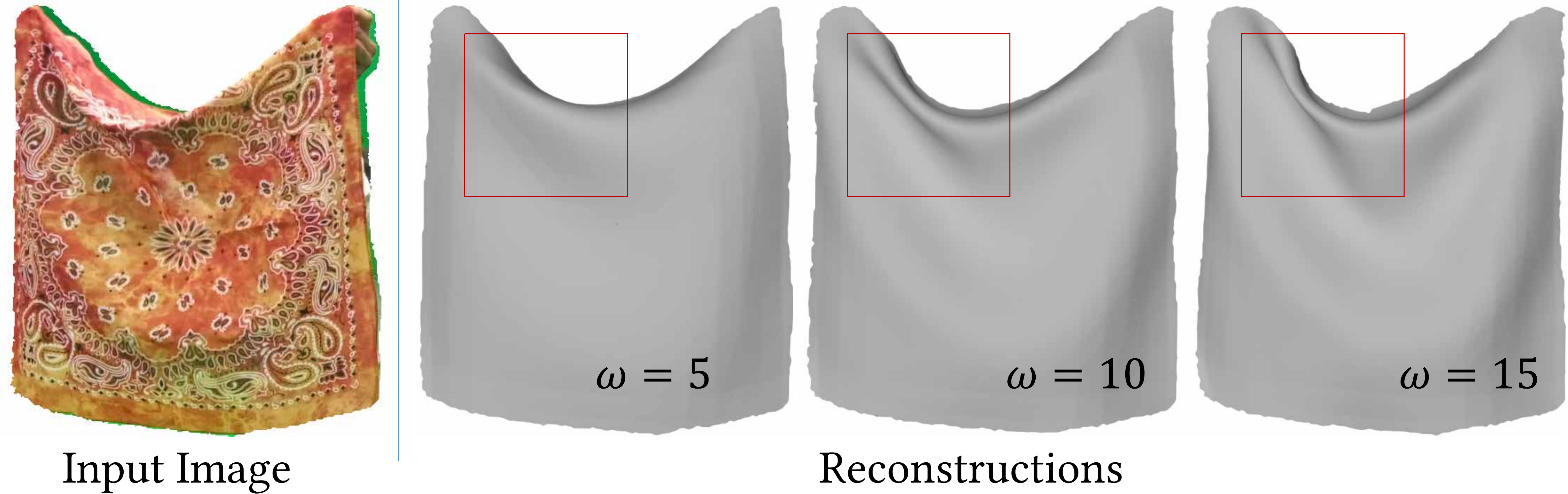}
   \caption{
   \textbf{Sharpness control.} %
   The amount of deformation details in the reconstructions can be tuned by varying NDF $\omega$. 
   }
   \label{fig:smoothness}
\end{figure}

%% file: figures/fig_gaussian_count.tex
\begin{figure}
  \centering
  \includegraphics[width=0.97\linewidth]{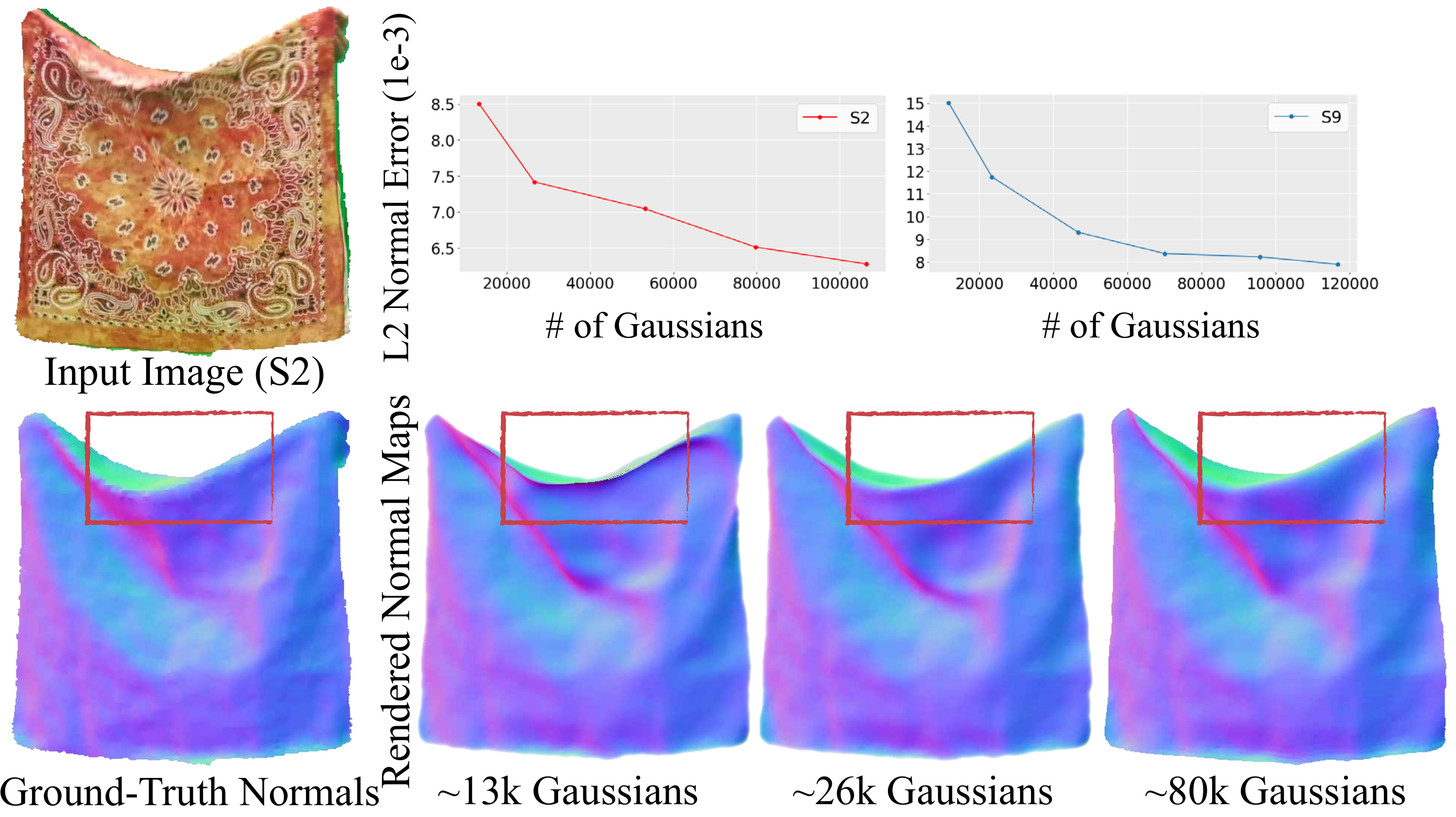}
   \caption{
  \textbf{Gaussian count.}
    (top:) Reconstruction error for the varying number of Gaussians ($N_g$); (bottom:) Even at lower $N_g$s, our method tracks surfaces with fine-grained details, although with slightly lesser accuracy.}
   \label{fig:gaussian_count}
\end{figure}

%% file: tables/tab_ablation_transposed.tex
\begin{table}[tb]
  \centering
\begin{adjustbox}{width=\linewidth,center}
\begin{tabular}{rrrrrrrrrrr}
   \toprule
    Seq & S1 & S2 & S3 & S4 & S5 & S6 & S7 & S8 & S9 & \textbf{Avg} \\
    \midrule
    w/o physics prior & 1.96 & 1.65 & 7.01 & 35.28 & 56.60 & 15.29 & 117.5 & 41.36 & 31.61 & \textbf{34.25} \\
    w/o surface Gaussians & 5.96 & 20.77 & 9.46 & 49.72 & 13.71 & 8.80 & 6.80 & 5.40 & 6.41 & \textbf{14.00} \\
    w/o momentum & \cellcolor{third} 1.52 & 1.00 & \cellcolor{third}3.67 & 6.02 & \cellcolor{second}8.94 & 3.15 & \cellcolor{third} 4.87 & \cellcolor{second} 2.25 & \cellcolor{third} 3.56 & \textbf{3.89} \\
    w/o normal scale & \cellcolor{second}1.20 & \cellcolor{first} 0.53 & \cellcolor{first}3.14 & \cellcolor{first} 5.40 & \cellcolor{first}8.69 &  \cellcolor{third} 3.07 & 5.14 & 2.30 & 4.29 & \cellcolor{third} \textbf{3.75} \\
    w/o mask loss & 1.61 & \cellcolor{third} 0.62 & \cellcolor{second} 3.49 & \cellcolor{third} 5.67 & \cellcolor{third} 9.05 & \cellcolor{second} 2.91 & \cellcolor{second} 4.24 & \cellcolor{first} 1.90 & \cellcolor{first} 2.26 & \cellcolor{second} \textbf{3.53} \\
    \midrule
    Ours (full)  & \cellcolor{first} 1.17 & \cellcolor{second} 0.55 &  \cellcolor{second} 3.49 & \cellcolor{second} 5.66 & \cellcolor{first} 8.69 & \cellcolor{first} 2.51 & \cellcolor{first} 3.80 & \cellcolor{third} 2.27 & \cellcolor{second} 3.00 & \cellcolor{first} \textbf{3.46} \\
    \bottomrule
  \end{tabular}
  \end{adjustbox}
    \caption{
    \textbf{Detailed ablations.}
    We report the Chamfer distance for all sequences of~\phisft~dataset when ablating various design choices: without the physics loss, %
  without surface-induced Gaussian parameters, 
  without fixing Gaussian scale along the surface normal, 
  without momentum regularisation, 
  and without mask loss.
  }
  
  \label{tab:ablation_transposed}
\end{table}

%% file: figures/fig_additional_results.tex
\begin{figure*}
        \includegraphics[width=1\linewidth]{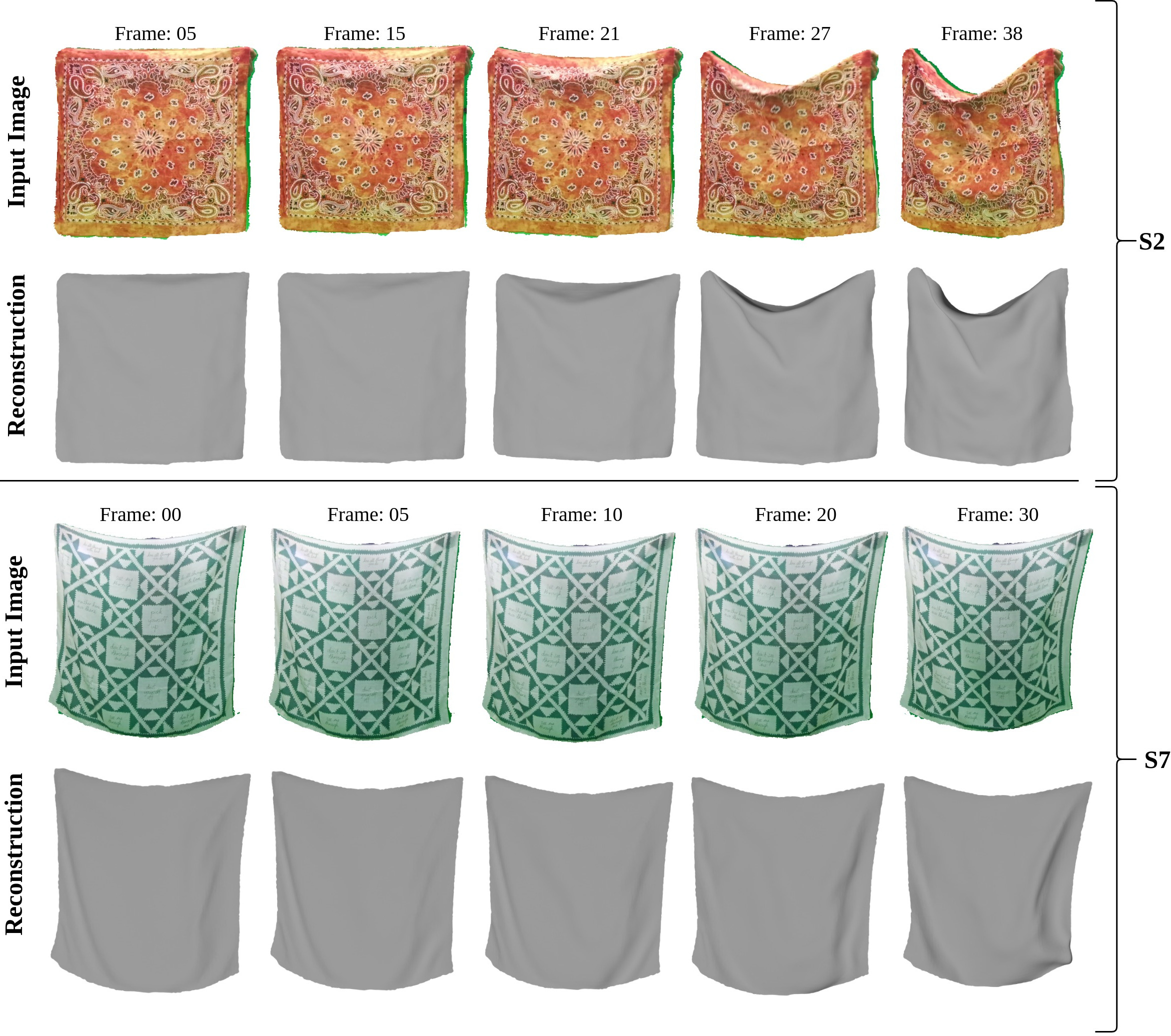} %
	\caption
	{
        \textbf{Examplary 4D surface tracking results by \ours.}
	We show additional qualitative results on two sequences. %
    Our method can reconstruct high-quality wrinkles and deformations just from the monocular video. 
    Please see the supplementary video for tracked reconstructions of all sequences.
        }
	\label{fig:suppl_result}
    \vspace{-10pt}
\end{figure*}

%% file: figures/fig_limitation.tex
\begin{figure}
  \centering
  \includegraphics[width=0.62\linewidth]{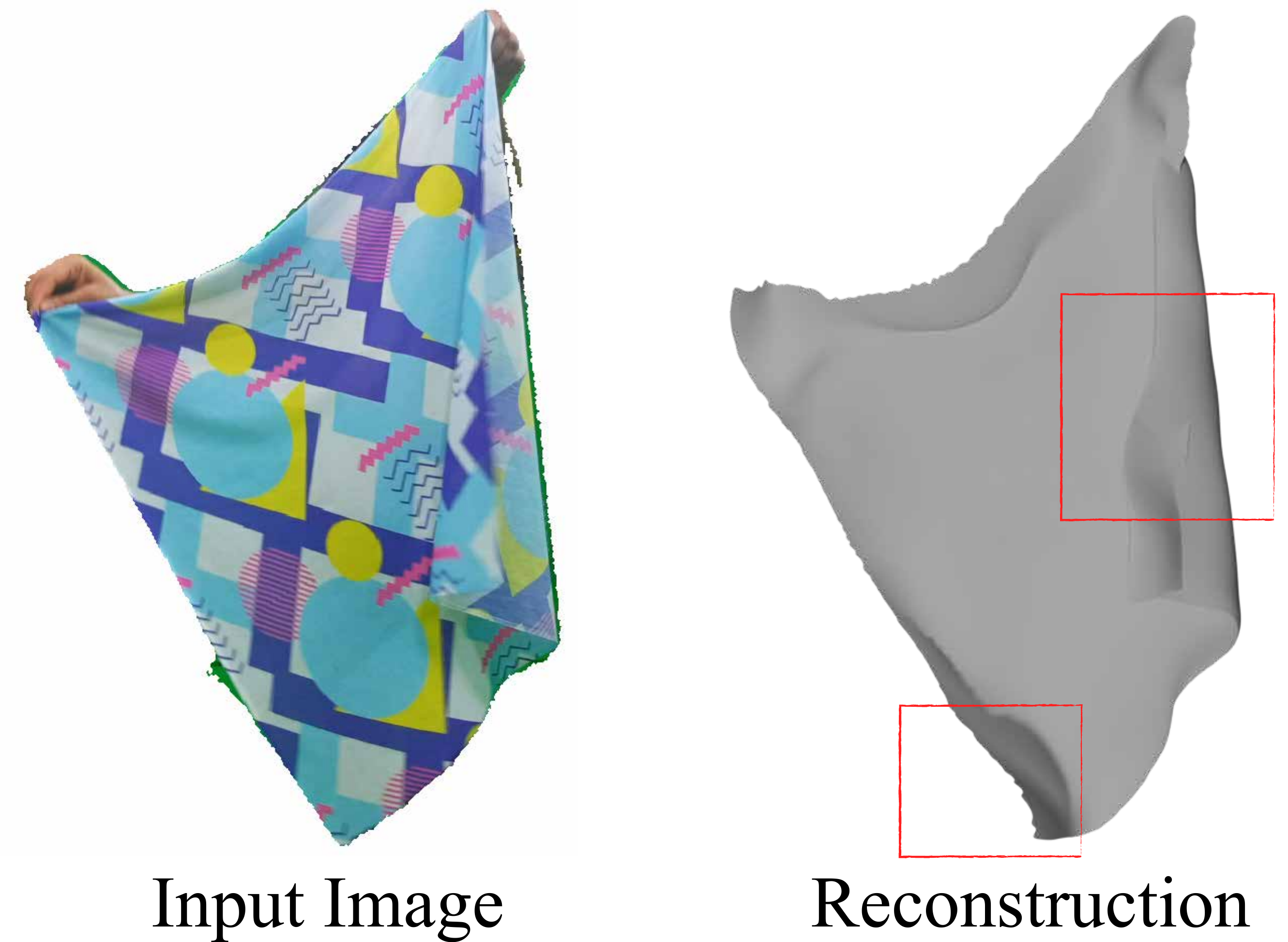}
   \caption{
   \textbf{Visualisation of the limitation.} %
   Our method does not handle the self-collision of tracked surfaces. Moreover, appearance changes due to deformation (\eg, shadows) can lead to minor artefacts. 
   }
   \label{fig:limitation}
   \vspace{-16pt}
\end{figure}